# On how religions could accidentally incite lies and violence: Folktales as a cultural transmitter

Quan-Hoang Vuong

Manh-Tung Ho

Hong-Kong T. Nguyen

Viet-Phuong La

Thu-Trang Vuong

Thi-Hanh Vu

Minh-Hoang Nguyen

Manh-Toan Ho

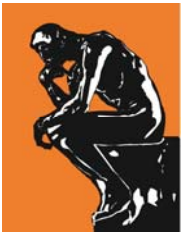

*A.I. for Social Data Lab*

**Tech:** QH Vuong 0000-0003-0790-1576; VP La 0000-0002-4301-9292
**Research team:** HM Tung 0000-0002-4432-9081; HKT Nguyen 0000-0003-4075-9823; TT Vuong 0000-0002-7262-9671; HM Toan 0000-0002-8292-0120
**Admin:** Thu-Ha Dam (Vuong & Associates, Hanoi, Vietnam)
**Email:** qvuong.ulb@gmail.com

# On how religions could accidentally incite lies and violence: Folktales as a cultural transmitter

Quan-Hoang Vuong [1,2,*]
Hong-Kong T. Nguyen [3]
Manh-Tung Ho [1,3,4]
Viet-Phuong La [1]
Thu-Trang Vuong [5]
Thi-Hanh Vu [6]
Minh-Hoang Nguyen [3]
Manh-Toan Ho [1]

[1] Center for Interdisciplinary Social Research, Phenikaa University, Ha Dong, Hanoi 100803, Vietnam

[2] Centre Emile Bernheim, Université Libre de Bruxelles, 1050 Bruxelles, Belgium

[3] Ritsumeikan Asia Pacific University, Oita Prefecture, Beppu City, Jumonjibaru, 1-1, 874-8577 Japan

[4] Institute of Philosophy, Vietnam Academy of Social Sciences, Hanoi, 59 Lang Ha, 100000 Vietnam

[5] Ecole doctorale, Sciences Po Paris, 75337 Paris, France

[6] Foreign Trade University, Hanoi 100000, Vietnam

[*] Corresponding author: qvuong@ulb.ac.be

**Abstract**

This research employs the Bayesian network modeling approach, and the Markov chain Monte Carlo technique, to learn about the role of lies and violence in teachings of major religions, using a unique dataset extracted from long-standing Vietnamese folktales. The results indicate that, although lying and violent acts augur negative consequences for those who commit them, their associations with core religious values diverge in the final outcome for the folktale characters. Lying that serves a religious mission of either Confucianism or Taoism (but not Buddhism) brings a positive outcome to a character ($β_{T\_and\_Lie\_O}$= 2.23; $β_{C\_and\_Lie\_O}$= 1.47; $β_{T\_and\_Lie\_O}$= 2.23). A violent act committed to serving Buddhist missions results in a happy ending for the committer ($β_{B\_and\_Viol\_O}$= 2.55). What is highlighted here is a glaring double standard in the interpretation and practice of the three teachings: the very virtuous outcomes being preached, whether that be compassion and meditation in Buddhism, societal order in Confucianism, or natural harmony in Taoism, appear to accommodate two universal vices—violence in Buddhism and lying in the latter two. These findings contribute to a host of studies aimed at making sense of contradictory human behaviors, adding the role of religious teachings in addition to cognition in belief maintenance and motivated reasoning in discounting counterargument.

**Keywords:** Buddhism; Confucianism; Taoism; violence; lies; double standard; folktales; Bayesian network modeling; 'bayesvl' R package

**Introduction**

Folklore materials offer one of the most imaginative windows into the livelihood and psychology of people from different walks of life at a certain time. These colorful narratives bring to life the identities, practices, values, and norms of a culture from a bygone era that may provide insights on speech play and tongue-twisters (Nikolić & Bakarić, 2016), habitat quality of farmers (Møller, Morelli, & Tryjanowski, 2017), treatments for jaundice (Thenmozhi et al., 2018), and contemporary attitudes and beliefs (Michalopoulos & Xue, 2019). While the stories tend to honor the value of hard work, honesty, benevolence, and many other desirable virtues, many of such messages are undercut by actions that seem outlandish, morally questionable, or brutally violent (Alcantud-Diaz, 2010, 2014; Chima & Helen, 2015; Haar, 2005; Meehan, 1994; Victor, 1990). In a popular Vietnamese folktale known as "Story of a bird named *bìm bịp* (coucal)," a robber who repents on his killing and cuts open his chest to offer his heart to the Buddha gets a better ending than a Buddhist monk who has been religiously chaste for his whole life but fails to honor his promise to the robber—i.e. bringing the robber's heart to the Buddha. In his quest for the robber's missing heart, not only does the monk never reach enlightenment, but he also turns into a coucal, a bird in the cuckoo family (Figure 1).

On the one hand, the gory details of this story likely serve to highlight the literal determination and commitment of the robber to repentance, which is in line with the Buddhist teaching of turning around regardless of whichever wrong directions one has taken. On the other hand, it is puzzling how oral storytelling and later handwriting traditions have kept alive the graphic details—the images of the robber killing himself in the name of Buddhism, a religion largely known for its non-violence and compassion. Aiming to make sense of these apparent contradictions, this study looks at the behavior of Vietnamese folk characters as influenced by long-standing cultural and religious factors. The focus on the folkloristic realm facilitates the discovery of behavioral patterns that may otherwise escape our usual intuitions.

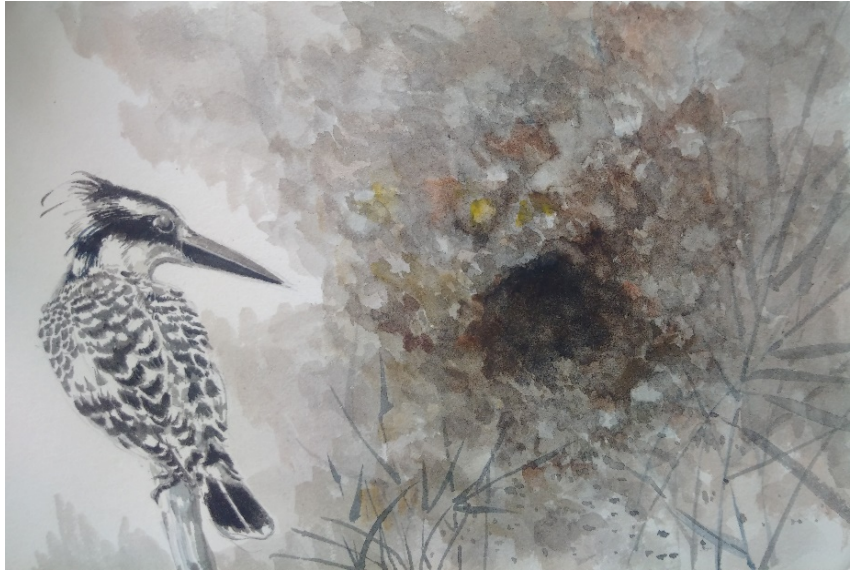

**Figure 1: "The Appalled Bird" by Vietnamese artist Bui Quang Khiem (watercolor, 2017)**

In order to highlight the unique interpretation of the possible interplay between religious teachings/values and deviant behaviors such as violence and lies in folklore, this study applies Bayesian networks analysis, which is based on conditional probability and helps researchers reduce the risk of overestimating effects or making logical inconsistencies (Downey, 2012; Gill, 2002; Jackman, 2000, 2009; Kruschke, 2015; Malakoff, 1999; McElreath, 2016). Indeed, while scholars have pointed out the prevalence of elements related to violence and lies in folktales (Alcantud-Diaz, 2010, 2014; Chima & Helen, 2015; Haar, 2005; Meehan, 1994; Victor, 1990), few have offered a rigorous statistical method to understand the interactions between these elements and their religio-cultural contexts. The research method follows the wave of studies on computational folkloristics, which emphasize the digitization of resources, the classification of folklore, and the necessary algorithms for data structure development (Abello, Broadwell, & Tangherlini, 2012; Dogra, 2018; Nguyen, Trieschnigg, & Theune, 2013; Tangherlini, 2013; Tehrani & d'Huy, 2017).

The scope of the present research, however, differs from the largely Euro-centric research projects due to its focus on Vietnamese folktales (Bortolini et al., 2017a, 2017b; d'Huy, Le Quellec, Berezkin, Lajoye, & Uther, 2017; Nguyen et al., 2013; Nikolić & Bakarić, 2016). It is, in fact, an expansion of an earlier project that examines the "cultural additivity" in Vietnamese culture – a phenomenon understood as the selection and inclusion of ideas, beliefs, or artefacts that may sometimes appear contradictory to principles of their existing beliefs to their culture (Vuong et al., 2018). Given that there is a certain degree of interactions among the elements constitutive of the three religions of Confucianism, Buddhism, and Taoism, it is reasonable to hypothesize that there may be some relationship between these religiously-imbued teachings and the universally-frowned upon acts of lying and violence. The three religions make a good case study because, despite their deep-rooted influence in Vietnam over centuries, the values they uphold such as benevolence, loyalty-fidelity, justice-righteousness, propriety, compassion, non-violence, and honesty have not completely deterred the acts of lying and violence (Vuong et al., 2018).

In this regard, the present study contributes to the wave of scholarship on non-WEIRD (Western, Educated, Industrialized, Rich, and Democratic) countries for shedding light on the little-known behavioral variability and contradictions in the folklore of a developing Asian country. We argue that, despite the universality of lies and violence across societies, their interactions with institutional religious teachings can generate a cultural variance in terms of outcome.

**Literature Review**

*The relationship between religions and lying or violence*

The acts of lying and violence represent deviances to the acceptable moral norms regardless of the cultural and religious settings. When examined through the binary religion–secular dimension, it is widely believed that religiosity, with its emphasis on being, loving, compassionate, honest, humble, and forgiving, should create changes reflecting such virtues in the behavior of the religious followers. This assumption is supported by the theory of cognitive dissonance: because religious people have an internal motivation to behave consistently with their beliefs, any behaviors that are not so would result in dissonance (Festinger, 1962; Perrin, 2000). Along this line of argument, research on the role of religion frequently draws on the work of Emile Durkheim, who recognizes religion as the prime source of social cohesion and moral enforcement.

Yet, not just the clergymen who have doubts about the constraining effects of religious faith but also scholars over the ages. To make sense of the relationship between religiosity and deviant behaviors, scholars from as far back as the 1960s have sought to measure how church membership or religious commitment could deter delinquent activities, though pieces of empirical evidence over the years remain inconclusive (Albrecht, Chadwick, & Alcorn, 1977; Hirschi & Stark, 1969; Rohrbaugh & Jessor, 1975; Tittle & Welch, 1983). In their influential study, Hirschi and Stark (1969) ask if the Christian punishment of hellfire for sinners can deter delinquent acts among the firm believers, and surprisingly find no connection between religiosity and juvenile delinquency. Subsequent studies tend to fall along two lines, either confirming the irrelevance of religion and deviance (Cochran & Akers, 1989; Tittle & Welch, 1983; Welch, Tittle, & Grasmick, 2006), or pointing out certain inhibiting effect of religiosity depending on the types of religious contexts (Benda, 2002; Corcoran, Pettinicchio, & Robbins, 2012; Evans, Cullen, Dunaway, & Burton Jr, 1995; Rohrbaugh & Jessor, 2017). Additional studies have looked at religious contexts beyond the WEIRD (Western, educated, industrialized, rich, democratic) countries such as in South Korea and China but also reached inconsistent results on the religiosity–deviance relationship (Wang & Jang, 2018; Yun & Lee, 2016).

Notably, in the research literature on the relationship between religious teachings/commitment and misconduct, the spotlight has largely been on Christianity and its punitive supernatural systems (Perrin, 2000). Although non-religious people's moral attitude and behaviors can be drawn from their experiences and interactions with religious others (Sumerau & Cragun, 2016), the formulation of moral identity is a complex process involving conceptualization of the self over different developmental stages (Wainryb & Pasupathi, 2015). As such, one need not define religion merely as a "belief in spiritual beings" but should include the in-between spaces of spirit and non-spirit (E. B. Tylor as cited in Day, Vincett, and Cotter (2016)). This definition gives room for studying the influence of semi-religious teachings or folk religion in countries where the word *religion* itself does not evoke the same sentiment or understanding. This is precisely where the current study fits in, as we will explain below.

### *The portrayal of lies and violence in folklore*

The acts of hurting or killing one another are common images in folklore and religious narratives around the world (Chima & Helen, 2015; Haar, 2005; Houben & van Kooij, 1999; Meehan, 1994; Victor, 1990). This is attributable to the role violence plays in human storytelling—as a story device, it gives voices to both the offenders and the victims (Sandberg, Tutenges, & Copes, 2015) as well as serves interactional and recreational purposes (Coupland & Jaworski, 2003). How people tell their stories, lying or being honest, not only reflects but also allows us to grasp the intertwining nature of values, identities, and cultures (Sandberg, 2014). For example, Victor (1990) shows that rumors about the satanic cult—which is rooted in the mythologized ancient blood ritual and Satan's conflict with God—often arise during a period of intense social stresses and cultural crisis. Similarly, the high amounts of violent terms and actions in the Grimm's fairy tales have been shown to be tied up with power and social status in the construction of the self (Alcantud-Diaz, 2010, 2014). In a different study, Haar (2005) analyzes a number of motifs in Chinese witch-hunt stories, such as the consumption of adult human body parts, children, and fetuses, to illustrate the force of the anti-Christian movements and the interplay of folkloric fears and political history in China. This finding is supported by Tian (2014) when looking at the Tianjin Missionary Case of 1870.

In contrast to the wealth of studies on violence in folklore, there is scant research on the act of lying and its implications. This is surprising given how prevalent lying is across folk cultures. Lying tales make up one category of its own within the folktales of Thailand (MacDonald & Vathanaprida, 1994). Research studies that touch on this topic are understandably centered around the themes of honesty/dishonesty and moral development in storytelling (Kim, Song, Lee, & Bach, 2018; MacDonald, 2013). In a rare approach that examines the semantics of lies, one study compares the function of lies in folktales to a prosthesis in the domain of discourse, such that the use of lie transforms the story system from horizontal to vertical, hence, action plan to meta-action (Towhidloo & Shairi, 2017).

This survey highlights a gap in the literature on the interactions of different religions or religious teachings with deviant behaviors such as lying and killing in folklore. Even beyond the folkloristic realm, findings also remain inconclusive on the relationships between lying/cheating and religion (Bruggeman & Hart, 1996; Mensah & Azila-Gbettor, 2018; Rettinger & Jordan, 2005) as well as between violence and religion (Atran, 2016; Blogowska, Lambert, & Saroglou, 2013; Henrich, Bauer, Cassar, Chytilová, & Purzycki, 2019; Purzycki & Gibson, 2011). The primary reason is perhaps, different cultures or social groups recognize and penalize different sets of moral values (Graham, Meindl, Beall, Johnson, & Zhang, 2016; Haidt & Graham, 2007; Haidt & Joseph, 2004; Henrich, Heine, & Norenzayan, 2010; McKay & Whitehouse, 2015). In other words, while all religions stress the need to cultivate virtues such as loyalty, reciprocity, honesty, and moderation, how these virtues are practiced in reality are not universal across cultures. What is equally noteworthy is how certain vices, e.g., lying and violence, are portrayed and tolerated in different parts of the world. An analysis of the cultural history of South Asian has revealed the development of arguments that seemingly rationalize violence, turning violence into non-violence over the course of millennia (Houben & van Kooij, 1999). One example the authors point out is the glorification of the gods and goddesses who have committed the most extreme forms of violence.

Thus, while extant research has not confirmed the relationship between religious teachings and lying and/or violence, the interplay of these two variables may be different and better understood when looking through folklore—a colorful window into folk psychology. This is where the current study fits in—through the case of Vietnam, it looks at the two universal acts of violence and lies in storytelling to shed light on the influence of traditional religions on folkloristic behaviors.

*The Three Teachings in Vietnam*

Before delving in further, it is important to note that the Vietnamese word *tôn giáo* is not equivalent to its English translation *religion*, which is derived from the Latin root *religio* meaning 'to bind together' or 'to reconnect' (Durkheim, 1897). The Vietnamese word has its origin from the Chinese word *zongjiao* (宗教), which was imported from Japan (*shukyo* 宗教) in the first decade of the twentieth century (Casadio, 2016, p. 45). The word, comprised of *zong* as in "divisional lineage" and *jiao* as in "teaching," encompasses the praxis and doctrine of religion (Casadio, 2016). The word religion in Vietnamese can be interpreted as a "way of life" (the Chinese *dao* 道) or "teaching" (教) (Tran, 2017). In practice, the Vietnamese popular religion involves ancestor and deity worshipping, exorcism, spirit-possession, etc. (Cleary, 1991; Kendall, 2011; Toan-Anh, 2005; Tran, 2017). For this reason, the present study uses "the Three Teachings" to avoid the religious connotations and to instead refer to their influence in both lifestyle and traditional philosophies.

The fundamental contents of the Three Teachings are presented in Table 1. The details of these religions can be found in Vuong et al. (2018).

**Table 1. A summary of key contents of Confucianism, Taoism, and Buddhism as known in Vietnam**

|  | Confucianism | Taoism | Buddhism |
|---|---|---|---|
| **Earliest presence in Vietnam** | From 111 B.C. until A.D. 938 under Chinese domination (Nguyen, 1998) | $2^{nd}$ century AD from China (Xu, 2002) | $1^{st}$ or $2^{nd}$ century AD from India (Nguyen, 2014; Nguyen, 1985; Nguyen, 1998; Nguyen, 1993; Nguyen, 2008) |
| **Peak development** | Neo-Confucianism grew from the $15^{th}$ century to its peak in the $19^{th}$ century during the Nguyen dynasty (Nguyen, 1998, p. 93) | From the $11^{th}$ to $15^{th}$ centuries, during the Ly and Tran dynasties (1010-1400) (Xu, 2002) | $11^{th}$ century during the Ly dynasty (Nguyen, 2008, p. 19) |
| **Core teachings** | Three Moral Bonds, Three Obediences, Five Cardinal Virtues, Four Virtues (one set for women and another set for general) | Letting the natural flow of life, searching for longevity and immortality, and spiritual healing, which gets mixed into Vietnamese popular religious beliefs (Tran, 2017, p. 13) | The Four Noble Truths and the Eight-fold Path |
| **Core concepts** | moral conduct (*đức* 德), benevolence (*nhân* 仁), loyalty (*trung* 忠), wisdom (*trí* 智), filial piety (*hiếu* 孝), chastity or purity (*tiết* 節), righteousness (*nghĩa* 義), propriety (*lễ* 禮), integrity/faithfulness (*tín* 信) | "effortless action" (*vô vi* or *wuwei* 无为), "spontaneity" (*tự nhiên* 自然) | Karma (*nghiệp*): means the spiritual principle of cause and effect. It determines the cycle of reincarnation. |

As summarized above, the Three Teachings share cultivation of moral character but differ in the process and its end goal. For Confucianism, the process centers around building harmonious relationships with other society members and sustaining the societal structures. For Taoism, the emphasis is instead on protecting one's relationship with nature, keeping the natural flow of life to the point that one may detach oneself entirely from society. For Buddhism, the key to enlightenment (*nirvana*) is to understand the nature of reality—that life is suffering because one is ignorant of the impermanent nature of things.

None of the Three Teachings explicitly forbid lying, though Buddhist teachings do hold "Do not kill" as its first precept.

**Materials and Method**

This paper analyzes the outcome associated with behaviors of lying and violence of the main characters in selected Vietnamese folktales as well as the association of the Three Teachings with said behaviors. First, we encode the details of 307 Vietnamese folk stories into binary variables.

- **Lie:** whether the main character lies.
- **Viol:** whether the main character employs violence
- **VB:** whether the main characters' behaviors express the value of Buddhism
- **VC:** whether the main characters' behaviors express the value of Confucianism
- **VT:** whether the main characters' behaviors express the value of Taoism
- **Int1:** whether there are interventions from the supernatural world
- **Int2:** whether there are interventions from the supernatural world
- **Out:** whether the outcome of a story is positive for its main characters

For further details of the coding system, see Vuong et al. (2018). For example, if the main character behaves according to the core values of Buddhism, "VB" equals 1; if this character lies, "Lie" equals 1; if he or she commits violent acts, "Viol" equals 1. The details of the stories concerning Confucianism and Taoism are encoded similarly (La & Vuong, 2019; Vuong et al., 2018). We are also interested in whether external intervention from either human ("Int1") or the supernatural ("Int2") might influence the story's outcome. These data points are coded as blue in Figure 1.

Because Bayes' theorem makes no assumption about the infinite amounts of posterior data, all observations are probabilistic depending on prior distributions and can be updated by conditioning on newly-observed data (Gill, 2002; Kruschke, 2015; McElreath, 2016). Notably, although Bayesian inference is one of the more controversial approaches to statistics (Gelman, 2008), Bayesian statistics seems to offer a solution for the problem of irreproducibility (Editorial, 2017; Kruschke, 2015; McElreath, 2016), it reflects the approach of "mathematics on top of common sense" the Bayesian approach represents through the ability to update belief in light of new evidence (Scales & Snieder, 1997). The approach is, thus, especially helpful the social sciences where there various conflicting research philosophies (Vuong, Ho, & La, 2019b).

To be more precise, the Bayesian approach helps formalize the use of background knowledge to make more realistic inferences about a certain problem. With multilevel or hierarchical modeling, this idea is taken to another level where simultaneous analyses of individual quantities are performed (Spiegelhalter, 2019). Past studies in psychological and ecological sciences have demonstrated this effectiveness and flexibility of multilevel modeling. For example, Doré and Bolger found the data on the impacts of stressful life events on well-being are best fit with a varying curve model rather than a varying slope or a varying intercept model, which shows a wide range of different trajectories in life satisfaction different people to show a wide surrounding a negative life event (Doré & Bolger, 2018). A seminal study by Vallerand shows a hierarchical model of extrinsic and intrinsic motivation not only generates a framework to organize the literature on the subject, but also new and testable hypotheses (Vallerand, 1997) (Holman & Walker, 2018).

The **'bayesvl'** R package is coded up to support the Bayesian hierarchical multilevel analysis in this paper. They **bayesvl** package contains approximately 3,000 lines of code, which combine the ability of R to generate eye-catching graphics and the ability to simulate data of the Monte Carlo Markov Chain (MCMC) method (CITE Github). **bayesvl** was created based on some of the seminal works in Bayesian statistics such as (Kruschke, 2015; McElreath, 2016; Muth, Oravecz, & Gabry, 2018; Scutari, 2010; Thao & Vuong, 2015). One can find a complete user guide at (La & Vuong, 2019).

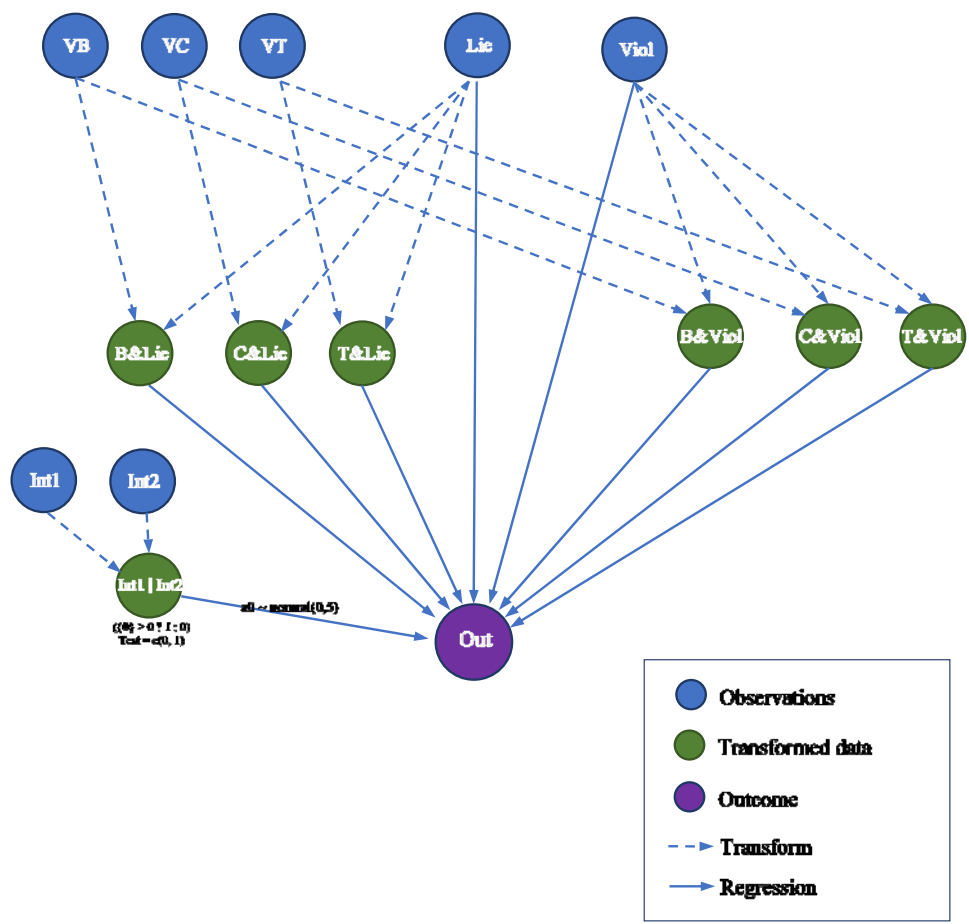

**Figure 2. The model of evaluating the influence of the Three Teachings ("VB", "VC", "VT") on lying ("Lie") and violent behavior ("Viol") of main characters in the folktales.**

The purpose of the model is to evaluate the influence of the *Three Teachings* ("VB", "VC", "VT") on lying ("Lie") and violent behavior ("Viol") of main characters in the folktales. This influence is evaluated based on whether the outcome of the stories is good or bad for the main character. Specifically, we are interested in finding out whether the main character lied or committed violent acts and at the same time, and their behaviors align with certain core values of the Three Teachings; for example, for simplicity's sake, the main character believes in the law of *karma* of Buddhism, yet lies and still succeeds in the end. With that in mind, we design a preliminary model, which is visually presented in Figure 1. We then perform the Bayesian MCMC analysis based on the multilevel model presented in Figure 1, in which, we measure the probability of an outcome of a character, given the religious values and the actions, to which he or she is committed.

In this study, we employed the Bayesian MCMC estimation uses 5000 iterations, 2000 warm-ups, and four chains; the results indicate a good fit of the model with data.

The analytical model, which is presented in Figure 1, can be coded using the following commands of the **bayesvl** R package.

```
# Design the model
model <- bayesvl()
model <- bvl_addNode(model, "O", "binom")
model <- bvl_addNode(model, "Lie", "binom")
model <- bvl_addNode(model, "Viol", "binom")
model <- bvl_addNode(model, "VB", "binom")
model <- bvl_addNode(model, "VC", "binom")
model <- bvl_addNode(model, "VT", "binom")
model <- bvl_addNode(model, "Int1", "binom")
model <- bvl_addNode(model, "Int2", "binom")

model <- bvl_addNode(model, "B_and_Viol", "trans")
model <- bvl_addNode(model, "C_and_Viol", "trans")
model <- bvl_addNode(model, "T_and_Viol", "trans")
model <- bvl_addArc(model, "VB",        "B_and_Viol", "*")
model <- bvl_addArc(model, "Viol",      "B_and_Viol", "*")
model <- bvl_addArc(model, "VC",        "C_and_Viol", "*")
model <- bvl_addArc(model, "Viol",      "C_and_Viol", "*")
model <- bvl_addArc(model, "VT",        "T_and_Viol", "*")
model <- bvl_addArc(model, "Viol",      "T_and_Viol", "*")
model <- bvl_addArc(model, "B_and_Viol", "O", "slope")
model <- bvl_addArc(model, "C_and_Viol", "O", "slope")
model <- bvl_addArc(model, "T_and_Viol", "O", "slope")

model <- bvl_addArc(model, "Viol",     "O", "slope")
```

Vuong et al. 2019. OSF Preprints, https://osf.io/nb7tg/

```
model <- bvl_addNode(model, "B_and_Lie", "trans")
model <- bvl_addNode(model, "C_and_Lie", "trans")
model <- bvl_addNode(model, "T_and_Lie", "trans")
model <- bvl_addArc(model, "VB",       "B_and_Lie", "*")
model <- bvl_addArc(model, "Lie",      "B_and_Lie", "*")
model <- bvl_addArc(model, "VC",       "C_and_Lie", "*")
model <- bvl_addArc(model, "Lie",      "C_and_Lie", "*")
model <- bvl_addArc(model, "VT",       "T_and_Lie", "*")
model <- bvl_addArc(model, "Lie",      "T_and_Lie", "*")
model <- bvl_addArc(model, "B_and_Lie", "O", "slope")
model <- bvl_addArc(model, "C_and_Lie", "O", "slope")
model <- bvl_addArc(model, "T_and_Lie", "O", "slope")

model <- bvl_addArc(model, "Lie",    "O", "slope")

model <- bvl_addNode(model, "Int1_or_Int2", "trans", fun = "({0}
> 0 ? 1 : 0)", out_type = "int", lower = 0)

model <- bvl_addArc(model, "Int1", "Int1_or_Int2", "+")
model <- bvl_addArc(model, "Int2", "Int1_or_Int2", "+")

model <- bvl_addArc(model, "Int1_or_Int2", "O", "varint", priors
= c("a0_ ~ normal(0,5)", "sigma_ ~ normal(0,5)"))
```

In the next section, we will go into the details of the model construction process.

*Interpreting the model*

First of all, Figure 1 is a logic map of the causal relationship between different (level) of variables and the outcome ("Out"), and this is a multi-level varying intercept model. Below is the most basic form of a multi-level varying intercept linear regression model.

$$y_i = \alpha_{j[i]} + \beta x_i + \epsilon_i$$

In the model, the variables "O", "Lie", "Viol",... are all binomially distributed, and they are presented by blue nodes in Figure 1 .

To evaluate the influence of the Three Teachings ("VB", "VC", "VT") and lying ("Lie") on the outcome of the stories, we join the Three Teachings variables with the lying variable to create transformed data. In Fig. 2, the transformed data are represented as green nodes , and the relation for transform data are represented in Figure 1 using the dash-line arrow ( ).

- **B_and_Lie**: the main character behaves according to the core values of Buddhism, yet there are details/contents in the story showing this character lies.

- **C_and_Lie**: the main character behaves according to the core values of Confucianism, yet there are details/contents in the story showing this character lies.
- **T_and_Lie**: the main character behaves according to the core values of Buddhism, yet there are details/contents in the story showing this character lies.

Similarly, we join the Three Teachings variables with the violent variable to create the transformed data below to evaluate the influence of the Three Teachings ("VB", "VC", "VT") and violent behavior ("Lie") on the outcome of the stories:

- **B_and_Viol**: the main character behaves according to the core values of Buddhism yet commits violent acts.
- **C_and_Viol**: the main character behaves according to the core values of Confucianism yet commits violent acts.
- **T_and_Viol**: the main character behaves according to the core values of Confucianism yet commits violent acts.

These variables are represented mathematically by the multiplication of the values of the observational data:

```
B_and_Lie = B * Lie
C_and_Lie = C * Lie
T_and_Lie = T * Lie

B_and_Viol = B * Viol
C_and_Viol = C * Viol
T_and_Viol = T * Viol
```

To evaluate whether the outcome of a story is changed because of intervention, whether from the supernatural or human, we combine the two observation variables "Int1" and "Int2" into one new transformed variable:

- **Int1_or_Int2**: there exists an intervention in a story of either the supernatural or the humans in the stories.

In terms of mathematical formalism, this variable is represented by a logical function below:

```
Int1_or_Int2 = (Int1 + Int2 > 0 ? 1 : 0)
```

To illustrate, the intervention of the supernatural ("Int1") can come from characters such as the Bodhisattva or the Buddha, or a fairy; the intervention of human ("Int2") comes from the people such as a king, a mandarine, or a landlord. As such, we have a new variable that combines the values of the two observational data "Int1" and "Int2," if this sum is greater than 0, it means at least one in two observation data takes the value of one (there is intervention). If the sum is equal to 0, the transformed data take on value 0; that means we have one variable to represent the intervention of either supernatural force (fairy, Buddha, etc.) or human force (king, mandarin, etc.).

We then move to plot the network again using R to double-check the logic of the model using the **bayesvl** R package:

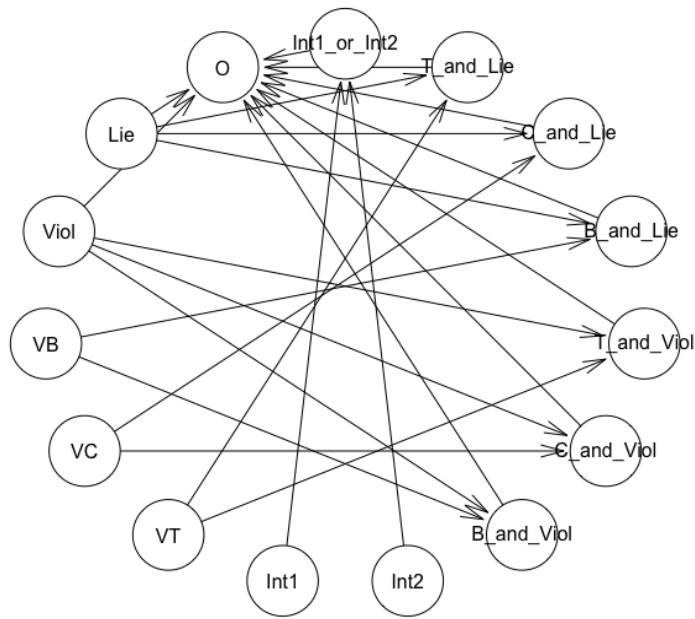

**Figure 3: Plotting the model using bayesvl R package.**

We have recreated the entire model in Figure 2 into a regression model in **bayesvl** (Figure 3). All Stan mathematical models and regressions will be created automatically accordingly. Moreover, the user can also check both the details of each node and also the entirety of the model, including all *nodes*, *transformed nodes*, and *logics* of each connection between two variables. To check the model specification, we use the following command.

```
> summary(model)

Model Info:
  nodes:     15
  arcs:      23
```

```
  scores:    NA
  formula:   O ~ b_B_and_Viol_O * VB*Viol + b_C_and_Viol_O * VC*Viol +
b_T_and_Viol_O * VT*Viol + b_Viol_O * Viol + b_B_and_Lie_O * VB*Lie +
b_C_and_Lie_O * VC*Lie + b_T_and_Lie_O * VT*Lie + b_Lie_O * Lie +
a_Int1_or_Int2[(Int1+Int2 > 0 ? 1 : 0)]

Estimates:
  model is not estimated!
```

**Checking conditional *posteriors*:**

The **bayesvl** package enables prediction of the outcome value of the model after regression. To execute the prediction, we need to add the test parameters when creating the nodes for the model. As can be seen, when creating node **Int1_or_Int2**, the **bayesvl** code has the following form:

```
model <- bvl_addNode(model, "Int1_or_Int2", "trans", fun = "({0}> 0 ? 1 
: 0)", out_type = "int", lower = 0, test = c(0, 1))
```

The paramter **test = c(0, 1)** allows **bayesvl** to add new codes to estimate "fixed predicted outcome" when **Int1_or_Int2 = 0** and **Int1_or_Int2=1**. This command will run a simulation for the model, the software will compute the sets of outcome value **yrep_Int1_or_Int2_1** and **yrep_Int1_or_Int2_2** after each regression iteration. Consequently, we will have *n* new value sets for the outcome.

*The priors of the model*

In the following box, the priors of the model can be double-checked.

```
> stan_priors(model)
    b_B_and_Viol_O ~ normal( 0, 10 );
    b_C_and_Viol_O ~ normal( 0, 10 );
    b_T_and_Viol_O ~ normal( 0, 10 );
    b_Viol_O ~ normal( 0, 10 );
    b_B_and_Lie_O ~ normal( 0, 10 );
    b_C_and_Lie_O ~ normal( 0, 10 );
    b_T_and_Lie_O ~ normal( 0, 10 );
    b_Lie_O ~ normal( 0, 10 );
    a0_Int1_or_Int2 ~  normal(0,5);
    sigma_Int1_or_Int2 ~  normal(0,5);
    u_Int1_or_Int2 ~ normal(0, sigma_Int1_or_Int2);
```

It is important to notice that most of the values of the *priors* here (and in real-life application) are set at default. Besides creating R/Stan statistical models based on a given logic map or checking the *priors*, **bayesvl** also allows the rechecking of the parameters using in a model through the function bvl_stanParams(model) or synthesize and simulate the data sample with the function **bvl_modelFit(model).**

## MCMC simulation

Markov Chain Monte Carlo (MCMC) is commonly used to simulate the probability distribution for the *posteriors* (Kruschke, 2015; McElreath, 2016; Spiegelhalter, 2019). The following command from **bayesvl** is to run the MCMC simulation in R:

```
> model <- bvl_addArc(model, "Int1_or_Int2", "O", "varint", priors = c("a0_ ~ 
normal(0,5)", "sigma_ ~ normal(0,5)"))
```

This command has four Markov chains to simulate the data sample. Each chain has 5000 iterations, in which, there are 2000 warm-up iterations, which means they cannot be counted into the effective sample size (**n_eff)**, the warm-up iterations are only for creating the stability of the chains.

**Analysis Results**

```
> summary(model)

Model Info:
  nodes:     15
  arcs:      23
  scores:    NA
  formula:   O ~ b_B_and_Viol_O * VB*Viol + b_C_and_Viol_O * VC*Viol +
b_T_and_Viol_O * VT*Viol + b_Viol_O * Viol + b_B_and_Lie_O * VB*Lie +
b_C_and_Lie_O * VC*Lie + b_T_and_Lie_O * VT*Lie + b_Lie_O * Lie +
a_Int1_or_Int2[(Int1+Int2 > 0 ? 1 : 0)]

Estimates:
Inference for Stan model: d4bbc50738c6da1b2c8e7cfedb604d80.
4 chains, each with iter=5000; warmup=2000; thin=1;
post-warmup draws per chain=3000, total post-warmup draws=12000.

                  mean se_mean   sd  2.5%   25%   50%   75% 97.5% n_eff Rhat
b_B_and_Viol_O    2.55    0.05 1.46  0.13  1.50  2.41  3.42  5.73   915 1.01
b_C_and_Viol_O   -0.28    0.01 0.61 -1.46 -0.68 -0.31  0.13  0.93  6689 1.00
b_T_and_Viol_O   -0.96    0.01 1.09 -3.21 -1.65 -0.91 -0.26  1.14  6820 1.00
b_Viol_O         -0.62    0.01 0.42 -1.43 -0.90 -0.62 -0.35  0.23  5892 1.00
b_B_and_Lie_O     0.70    0.02 1.44 -1.78 -0.28  0.56  1.52  4.03  6546 1.00
b_C_and_Lie_O     1.47    0.02 0.68  0.21  0.97  1.45  1.94  2.86  1676 1.01
b_T_and_Lie_O     2.23    0.02 1.59 -0.41  1.10  2.06  3.16  5.85  4523 1.00
b_Lie_O          -1.05    0.01 0.37 -1.77 -1.30 -1.05 -0.81 -0.32  3984 1.00
a_Int1_or_Int2[1] 1.20    0.00 0.21  0.78  1.05  1.20  1.33  1.62  7767 1.00
a_Int1_or_Int2[2] 1.35    0.00 0.19  0.99  1.23  1.35  1.48  1.73  3512 1.00
```

```
a0_Int1_or_Int2      1.18    0.04 1.34 -1.91 0.87 1.25 1.57 3.83 1353 1.00
sigma_Int1_or_Int2   1.49    0.04 1.82  0.04 0.28 0.78 1.98 6.67 1759 1.00
```

As a result, the model shows a good convergence, which is represented by two standard diagnostics of MCMC, **n_eff** (*effective sample size*) and **Rhat**. The values of **n_eff** show how many iterations of the Markov chain are needed for effective independent samples (McElreath, 2016). While the values of **Rhat** represents a more complicated simulation of the Markov chains converging toward a target distribution.

Normally, when **Rhat** is approximately 1, it means all chains have the same distribution; when **Rhat** greater than 1.1, it means the model has not converged; therefore, the samples are not credible. Meanwhile, it is a good signal for Bayesian inference when **n_eff** is above 1000 samples. In the current model, the results are good because most of the **Rhat**'s values are 1 and **n_eff** is more than 2000.

*Visualization and technical validation*

For Bayesian statistics, to evaluate the credibility of the analysis' results, one must carry out a process called visual diagnostics (La & Vuong, 2019; Muth et al., 2018; Vuong et al., 2019a; Vuong et al., 2018). **bayesvl** enables us to produce a graphic representation of the MCMC simulation results through the function **bvl_plotX,** in which, **X** represents the results for which we need to create visualizations, for example, the "Gelman shrink factors" (**Gelman**), the parameters (**Params**), or the pair parameters (**Pairs**).

*Markov chains visual diagnostics*

One can use **bvl_trace(model)** to generate the graphic representation of the MCMC chains. In our analysis, the chains are presented in Figure 4.

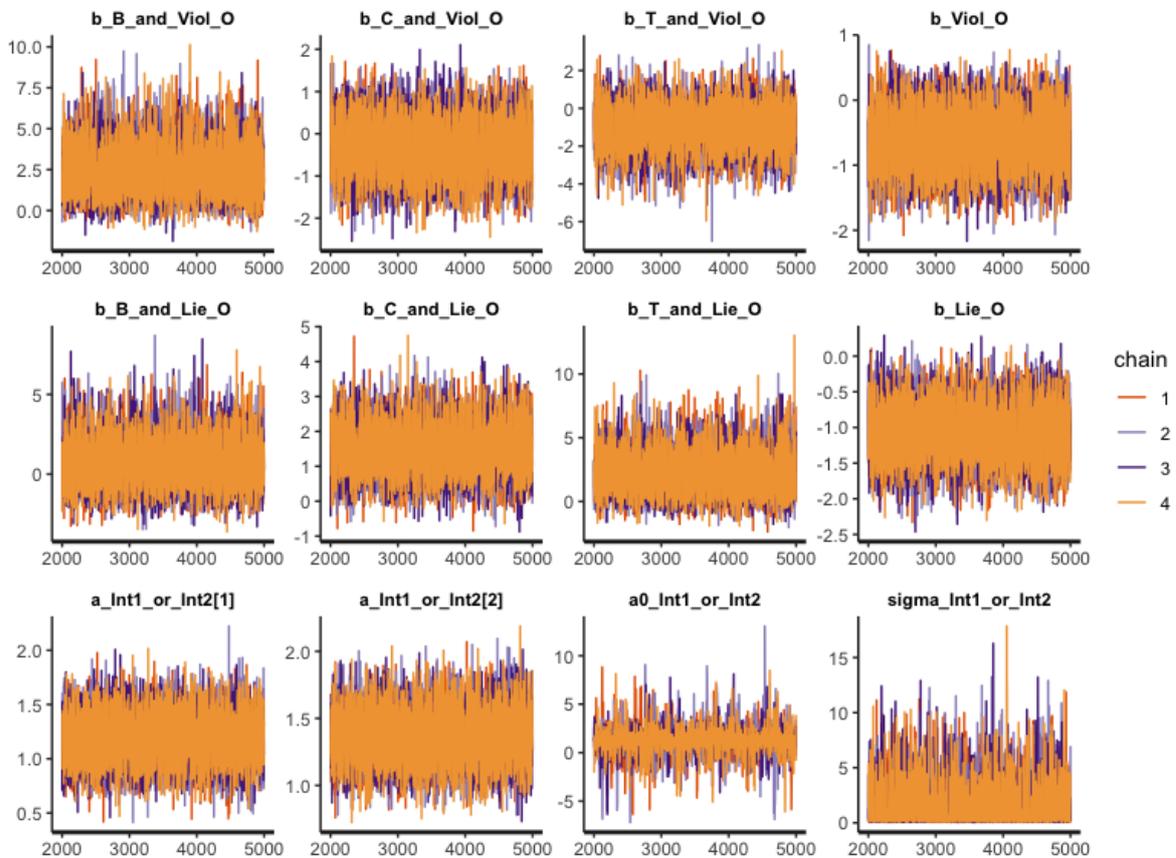

**Figure 4. The visual representation of the MCMC chains**

Each chain in Figure 4 has four component chains, each of which has 5000 *iterations*. Overall, there are no *divergent chains*—a strong signal for the autocorrelation phenomenon, which reflects the Markov property of the distribution. If we were to imagine that each chain has its own images, all of those images would be similar to each other. We can check the "Gelman shrink factor" in R using the following plot function of **bayesvl**:

```
> bvl_plotGelmans(model, NULL, 4, 3)
```

Gelman shrink factor plots can be checked through the graphics provided in Fig. 5.

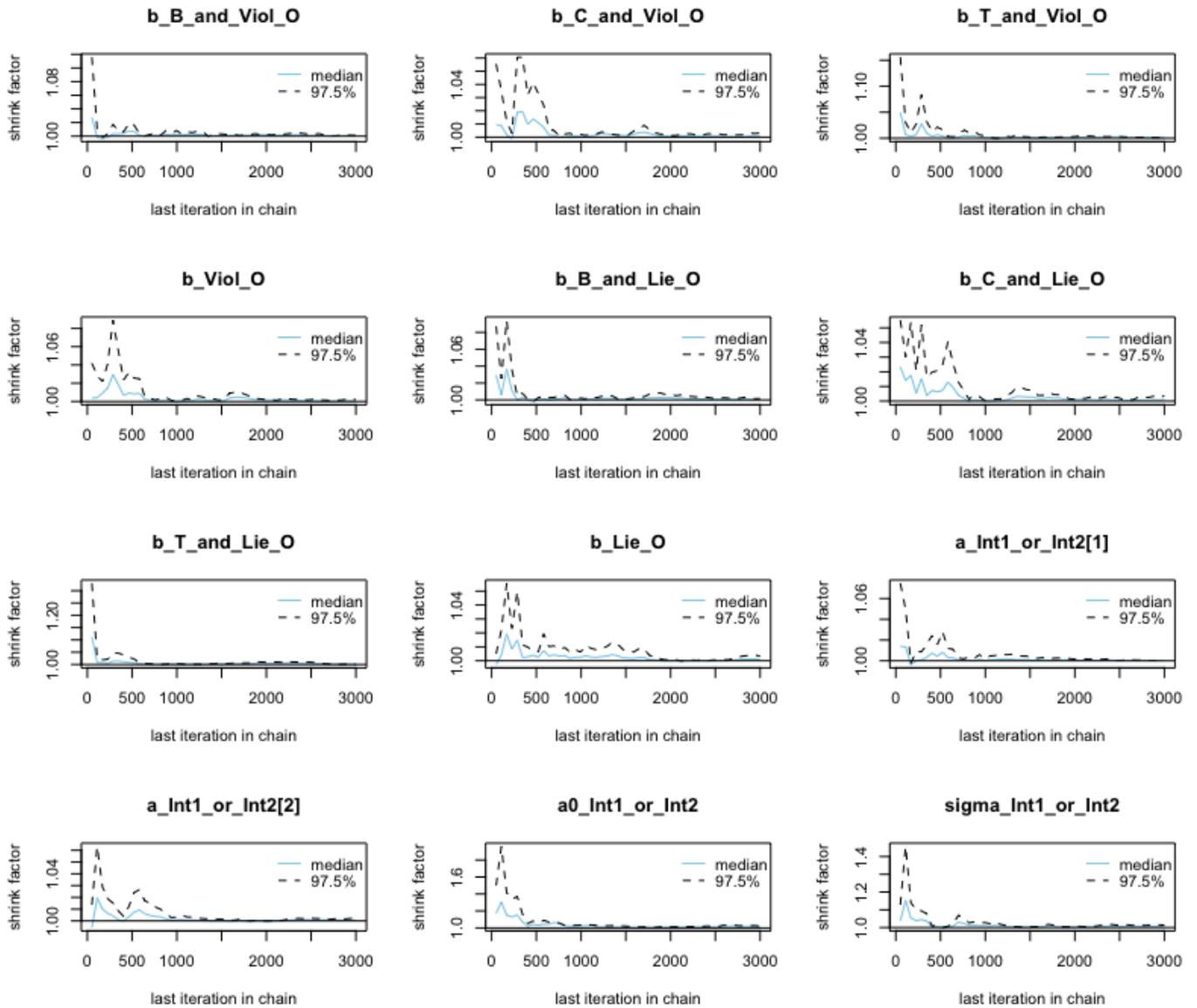

**Figure 5. The visual representation of the Gelman Shrink Factor.**

We find that the mean value of the potential scale reduction factor is 97.5%. Moreover, we also have the multivariate potential scale reduction, which Gelman and Brooks suggested (Brooks & Gelman, 1998). Figure 5 shows that the shrink factor converges to 1.0 quite rapidly, which satisfies the standards of MCMC simulation.

*Autocorrelation of each coefficient*

The MCMC algorithm produces the autocorrelated samples, not the independent samples. Therefore, the slow mixing due to too high acceptance rate or too low might lead to the process not ensure the Markov property. This check is to ensure after certain finite steps; autocorrelation will be eliminated (to 0). The plot function of the **bayesvl** R package for generating the graphic representation of the autocorrelation function (ACF) follows:

```
> bvl_plotAcfs(model, NULL, 4, 3)
```

The function helps to produce ACF graphs presented in Fig. 6.

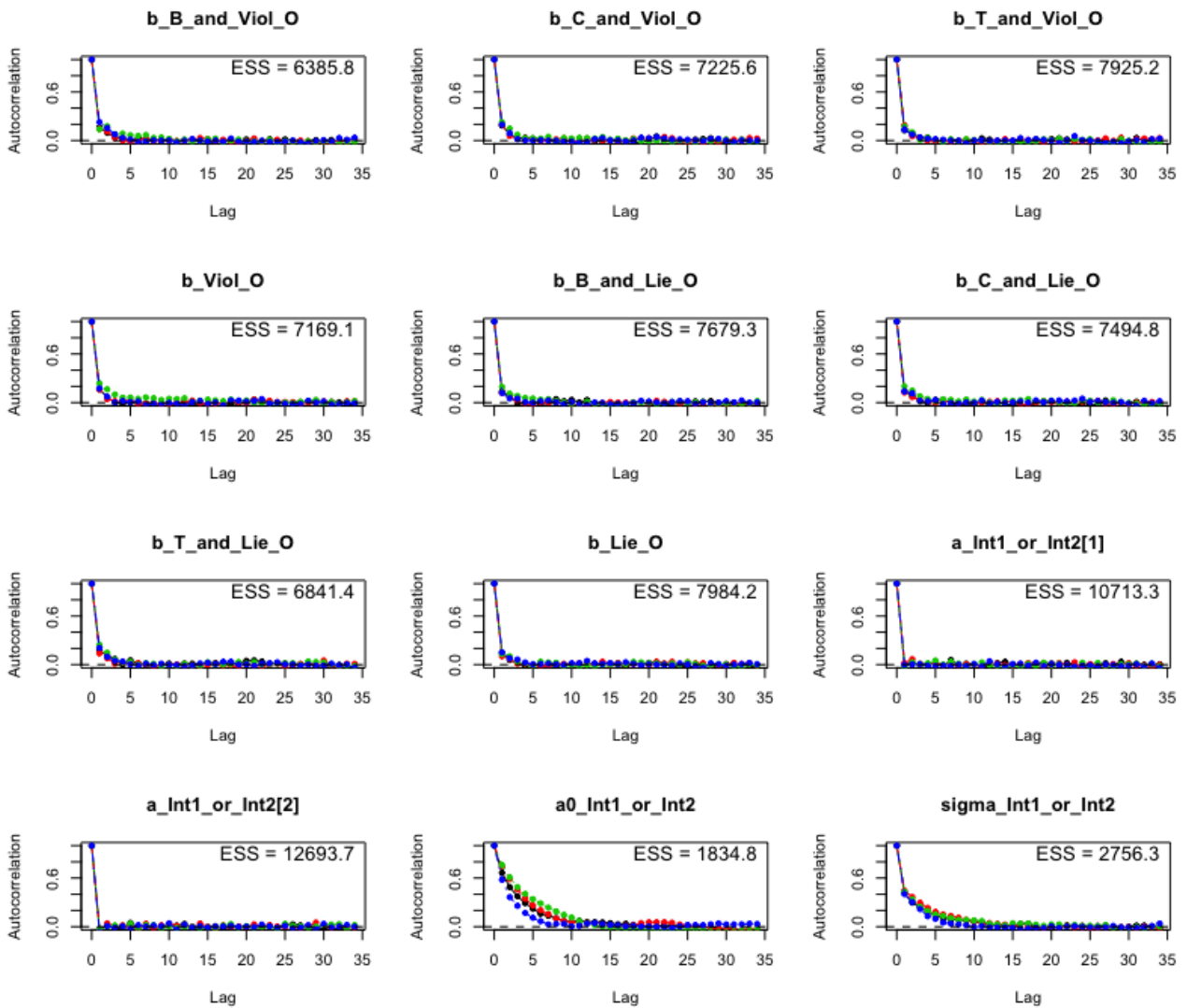

**Figure 6. Visual representation of the autocorrelation coefficient.**

Figure 6 shows that the effective sample sizes (ESS) for all coefficients are well above 1000. Besides, the coefficients most quickly converge before lag 3; the fact that lends support to computing efficiency and the Markov property of the chains.

*Assessing the regression coefficients*

In R, to compare the regression coefficients through the graphics, we have the following command of **bayesvl**,

```
> bvl_plotIntervals(model)
```

which helps produce the coefficients plot in Fig. 6.

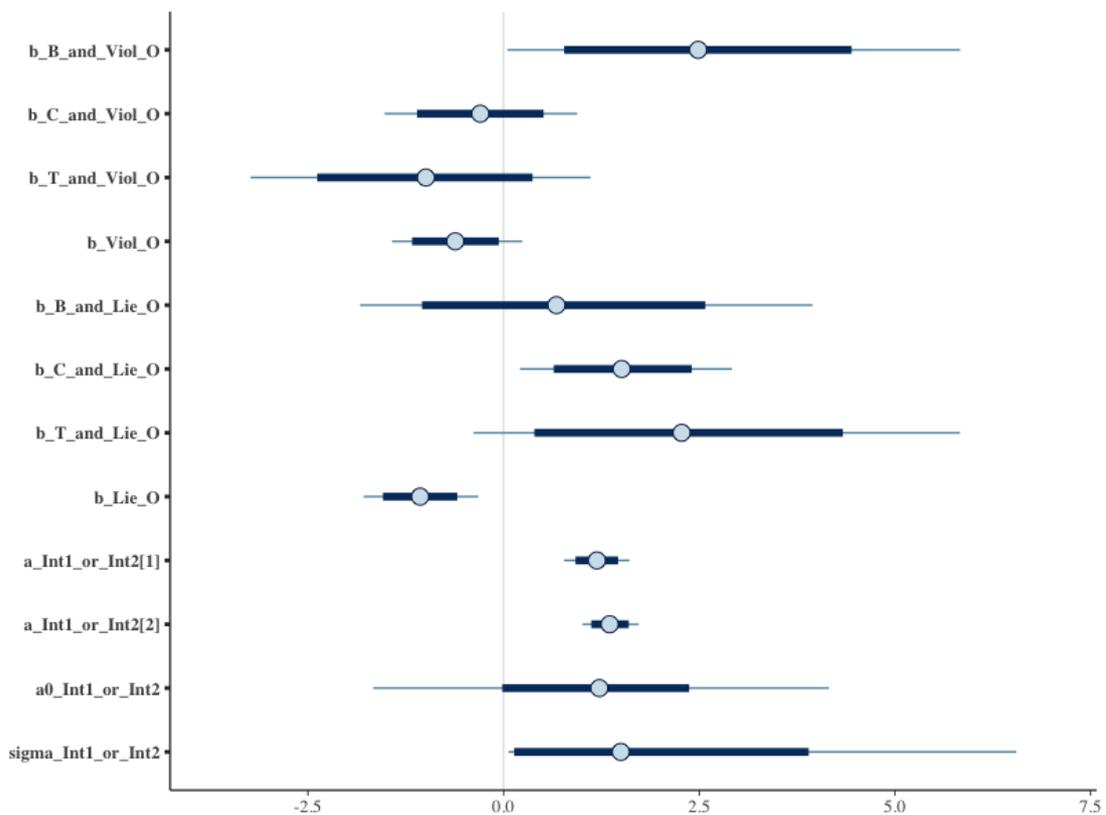

**Figure 7. Comparing the posterior distribution of the regression coefficients**

Alternatively, we can represent the distribution of regression coefficients visually is through the *Highest Posterior Distribution Intervals* (HPDI) using the following command:

```
> bvl_plotParams(model, 3, 3)
```

The distribution of all coefficients satisfies the technical requirements with HPDI (*Highest Posterior Distribution Intervals*) at 89% (Figure 7).

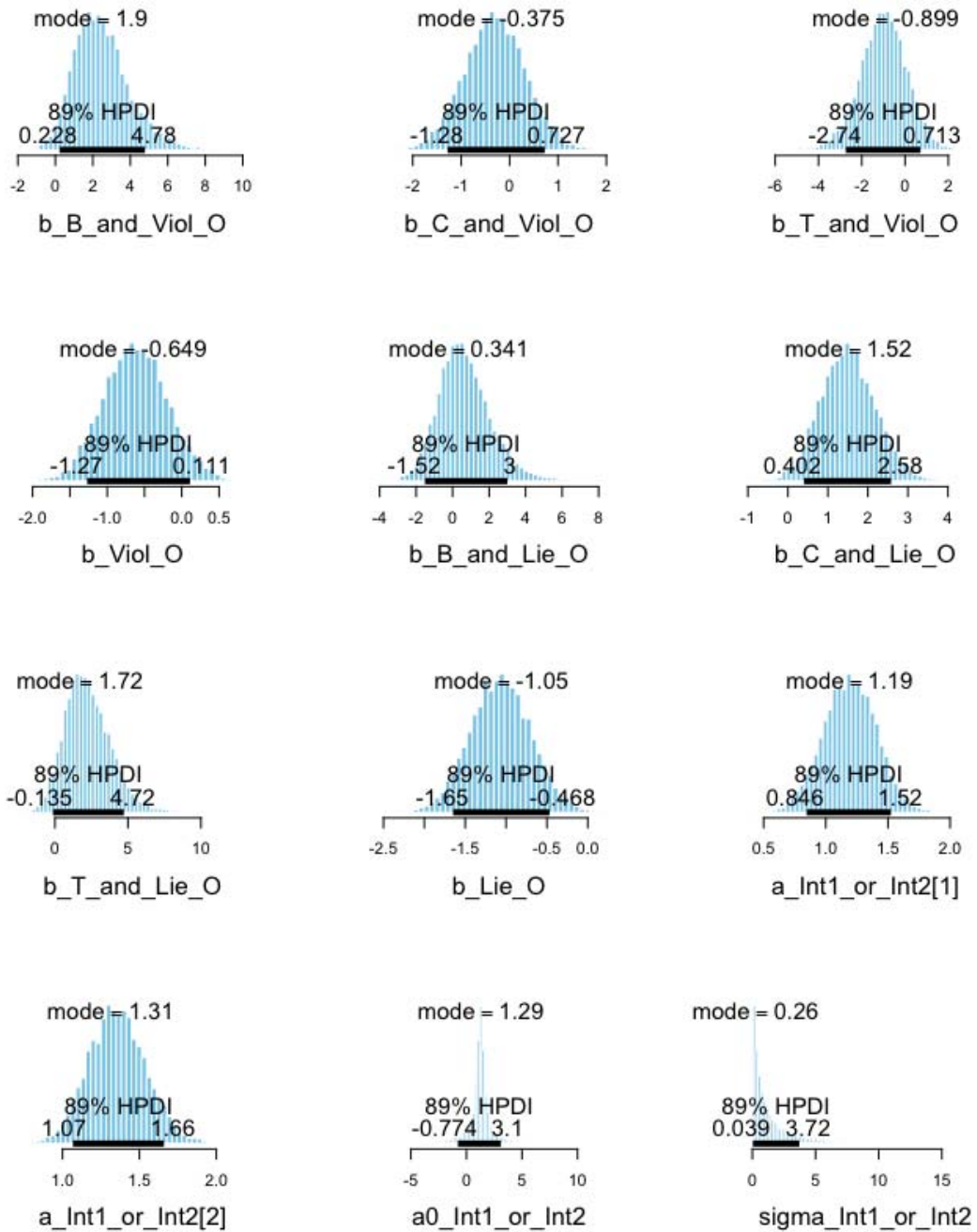

Vuong et al. 2019. OSF Preprints, https://osf.io/nb7tg/

**Figure 8. All coefficients satisfy the technical requirements with an 89% HPDI.**

From Figure 6 and Figure 7, when we only consider the outcome of lying and violence alone, the results are intuitive. The value of coefficients b_Lie_O (the probability of a good outcome given the main character lies) and b_Viol_O (the probability of a good outcome given the main character commits violent acts) are negative (-1.05 and -0.62, respectively). By contrast, when considering the interplay with religious values, it is not as straightforward to predict. The coefficients b_C_and_Viol_O and b_T_and_Viol_O are negative (-0.28 and -0.96, respectively) but b_B_and_Viol_O is positive (2.55). The figures indicate that the interaction among Confucian, Taoist values and violent acts bring about bad outcomes for the main characters, whereas for the main characters whose actions represent Buddhist values, violence tends to bring a good outcome. For lying, interaction with all three teachings tends to bring about good outcome for the main character ($\beta_{T\_and\_Lie\_O}$= 2.23; $\beta_{C\_and\_Lie\_O}$= 1.47; $\beta_{B\_and\_Lie\_O}$= 0.7).

***Assessing only the coefficients of the variables involving lying***

The bayesvl R command for this assessment follows:

```
> bvl_plotIntervals(model, c("b_B_and_Lie_O", "b_C_and_Lie_O",
"b_T_and_Lie_O", "b_Lie_O"))
```

Here are the images from analyzing the coefficients involved "Lie":

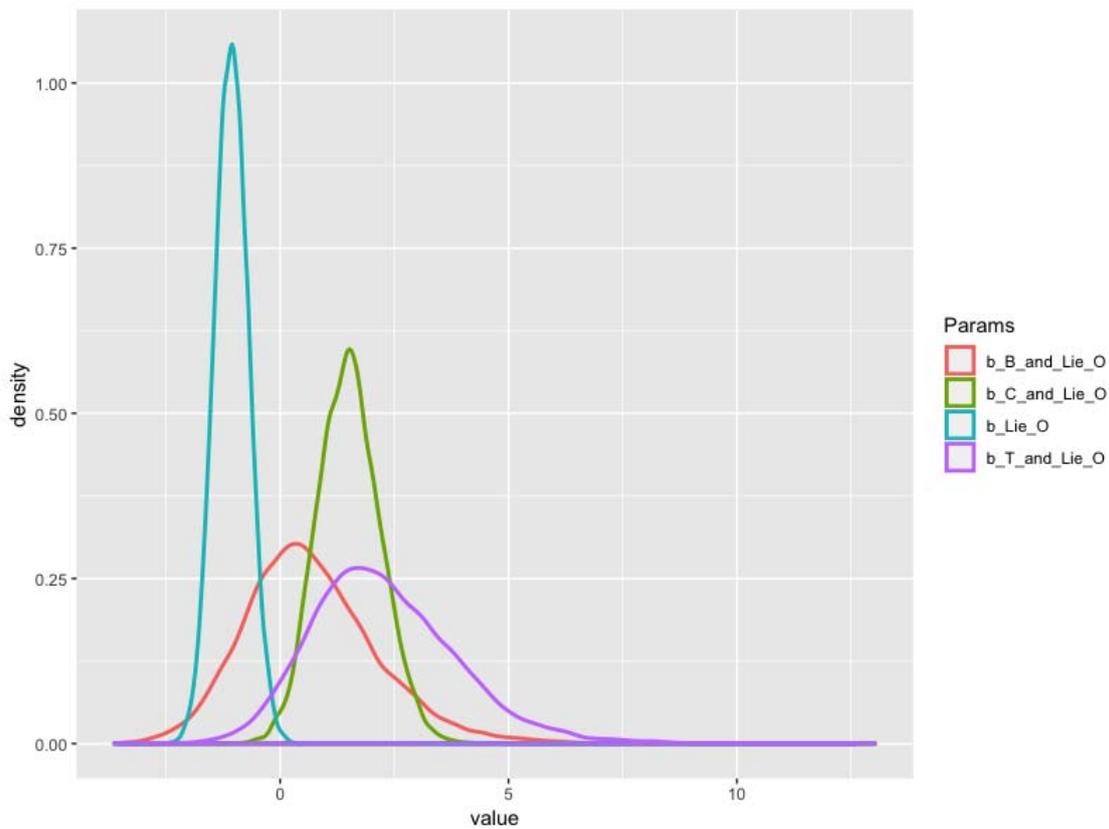

**Figure 9. Distribution of coefficients of variables involving lying**

The results in Figure 9 show that lying does not bring about good outcomes for the main character. The coefficient of **b_Lie_O** is negative, which indicates lying is associated with bad outcome for the folktale characters. This distribution is narrow, with a good credibility range.

However, as for the transformed data that involve the main character lying, and they express the core values of the Three Teachings, their coefficients are all positive. When there is the influence of Taoism, it seems likely that the main character enjoys a good outcome, though he or she might lie. Among the three religions, Buddhism is the least encouraging of lying, with **b_B_and_Lie_O** much smaller than the rest of the coefficients.

***Assessing only the variables involving violent actions***

```
> bvl_plotIntervals(model, c("b_B_and_Viol_O", "b_C_and_Viol_O",
"b_T_and_Viol_O", "b_Viol_O"))
```

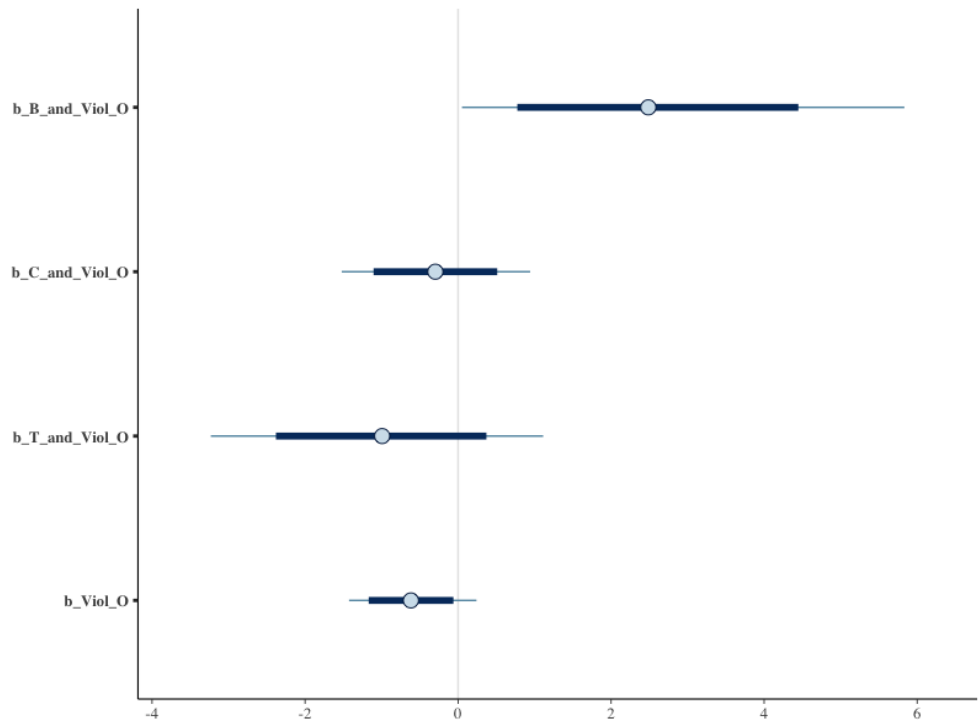

**Figure 10. Distribution of coefficients of variables involving violent acts.**

The R command to produce the image for the density **bvl_plotDensity**:

```
> bvl_plotDensity(model, c("b_B_and_Viol_O", "b_C_and_Viol_O",
"b_T_and_Viol_O", "b_Viol_O"))
```

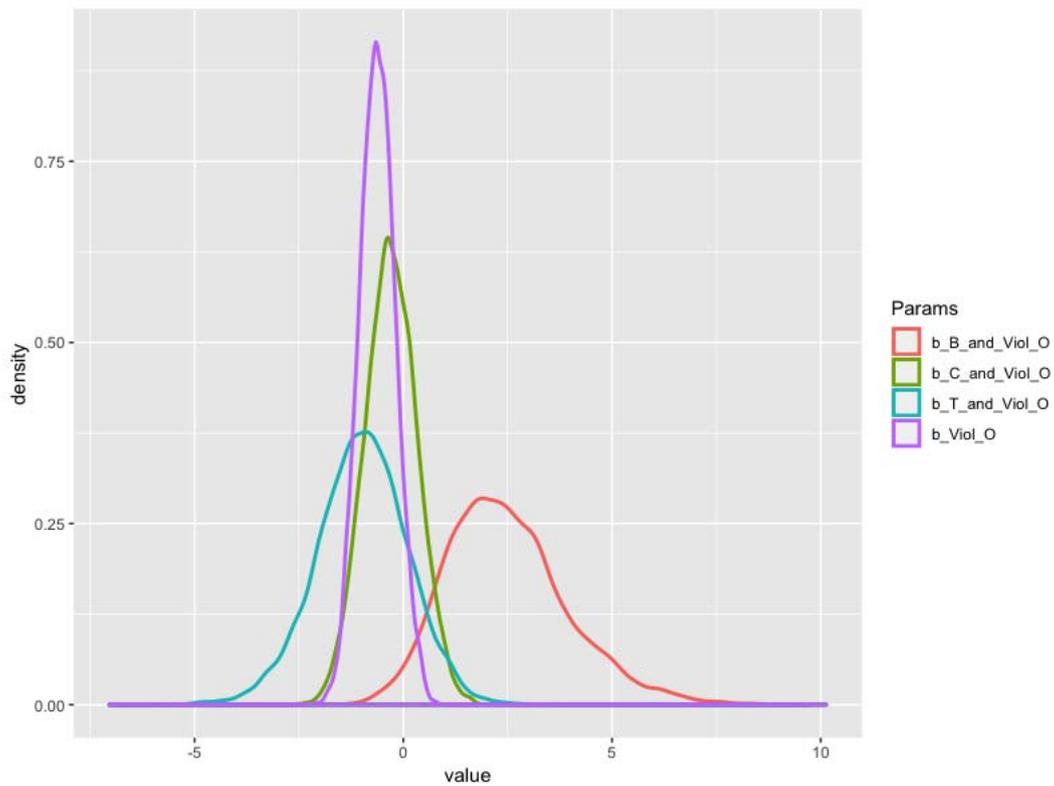

**Figure 11. Distribution of coefficients of variables involving violence.**

Overall, the results indicate that violence is not encouraged in the stories, as violence often bring bad outcomes for the main characters. The coefficient **b_Viol_O** is negative, which suggests violence tends to go against the good outcome of the main characters. This distribution is narrow, and with a good credibility range. When considering violence together with the Three Teachings variable, we can see that the coefficients for Confucianism and Taoism are still negative, but the coefficient for Buddhism and violence is positive. The result means that although the main character can commit a violent act, his or her outcome can still be good (positive coefficient).

*Comparing violence and lying*

```
> bvl_plotDensity2d(model, "b_Lie_O","b_Viol_O")
```

Both coefficients have the values of their simulated posteriors mostly in the negative-value region. They indicate characters who have lied and committed violence tend to have no good outcome.

Vuong et al. 2019. OSF Preprints, https://osf.io/nb7tg/

If we compare the correlation between this pair of parameters, the coefficient for violence is much weaker than that of violence. To compare the coefficients of Three Teachings elements with violence with each other, we look at Fig. 12.

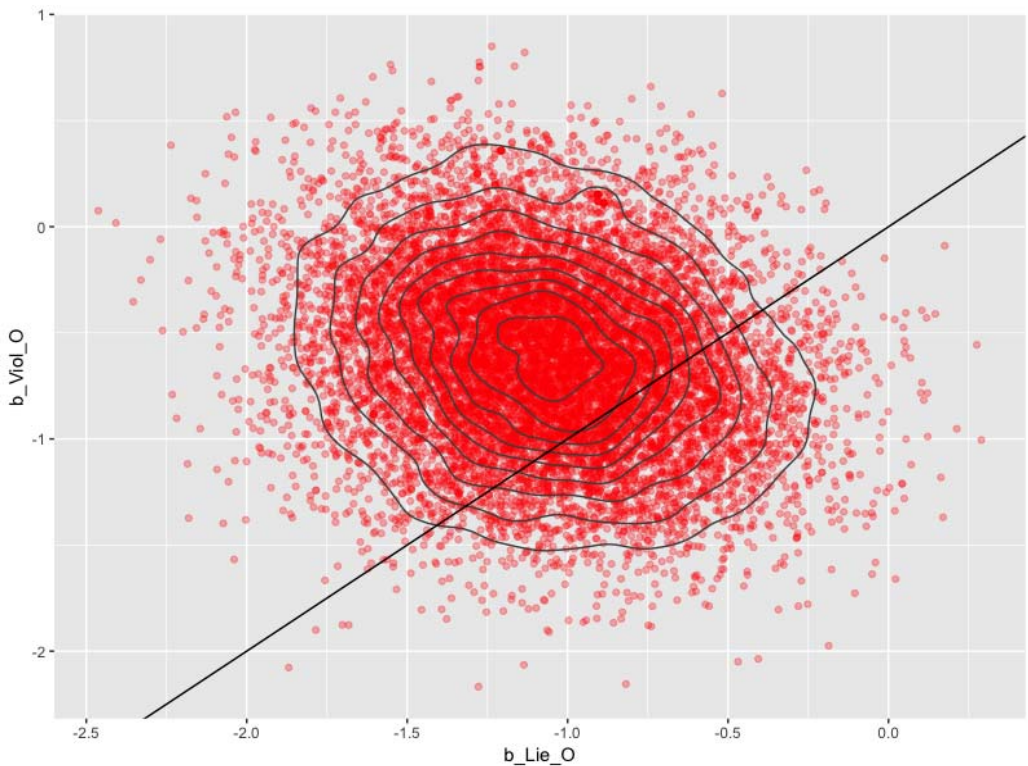

**Figure 12. Pair parameters comparison for b_Lie_O and b_Viol_O**

Both coefficients are negative, which indicates characters who have lied and committed violence tend to have no good outcome. If we compare the correlation of this pair of parameters, the coefficient for violence takes on smaller value than that of lying.

```
> bvl_plotDensity2d(model, "b_B_and_Viol_O", "b_C_and_Viol_O",
color_scheme = "orange")
```

We next examine the sample of simulated values of coefficient pairs ("b_B_and_Viol_O", "b_C_and_Viol_O") in Fig. 13.

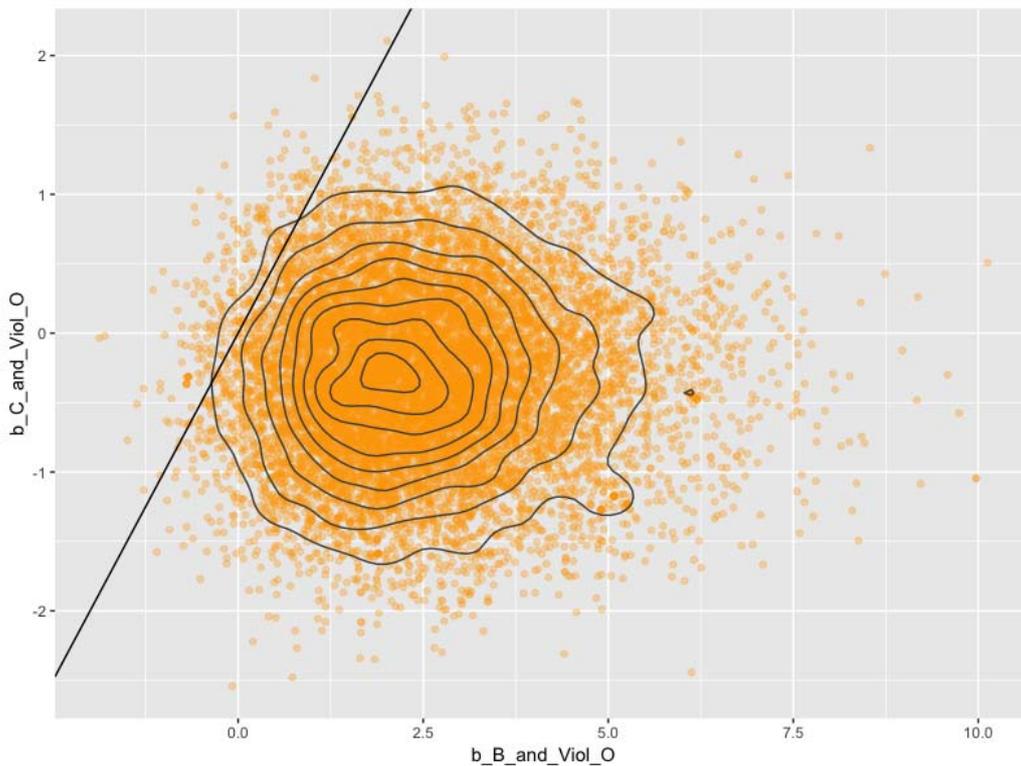

**Figure 13. Pair parameters comparison b_B_and_Viol_O and b_C_and_Viol_O**

The results indicate opposite trends for Buddhism and Confucianism. It seems the outcome tends to be good for characters that express the core values of Buddhism yet commit violence. Nonetheless, the influence of Buddhism is much stronger.

```
> bvl_plotDensity2d(model, "b_B_and_Viol_O", "b_T_and_Viol_O",
color_scheme = "orange")
```

Similarly, there are opposite trends for Buddhism and Taoism. It seems Buddhism is more tolerant toward violent behaviors, while it is the opposite for Confucianism. Nonetheless, the influence of Buddhism is much stronger. Next, we look at the coefficients of Confucianism and Taoism, which are plotted using the bayesvl density plot function.

```
> bvl_plotDensity2d(model, "b_C_and_Viol_O", "b_T_and_Viol_O",
color_scheme = "blue")
```

The function produces the graph in Fig. 14.

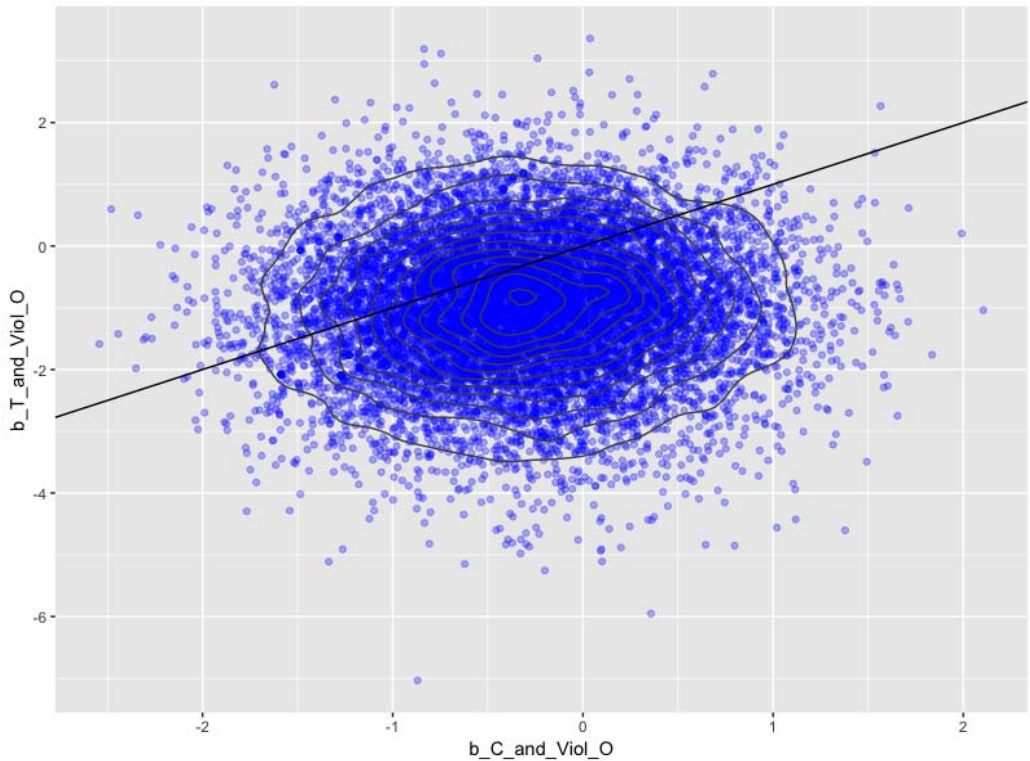

**Figure 14. Pair parameters comparison b_C_and_Viol_O and b_T_and_Viol_O**

In Figure 14, Confucianism and Taoism's coefficients are quite similar; they are both small negatives, centered around the same range of value. One can infer Confucianism and Taoism do not tolerate violence. The results, as shown in Figures 14 and 15, indicate a propensity to tolerate violent from characters whose values are Buddhism. The propensity contradicts with Buddhism's values, which encourage humanity and compassion. In terms of Confucianism and Taoism, we can reluctantly interpret the results based on the values that Confucianism and Taoism characters hold. Confucianism characters are usually scholars, while Taoism characters value the ethics of "non-contrivance" or "effortless action." Thus, the characters will tend to avoid using violence to reach their goals.

***Compare the correlations between the Three Teachings and lying behavior:***

First is the pair of Buddhism and Confucianism:

```
> bvl_plotDensity2d(model, "b_B_and_Lie_O", "b_C_and_Lie_O", color_scheme = "orange")
```

We have the result as in **Figure 15**.

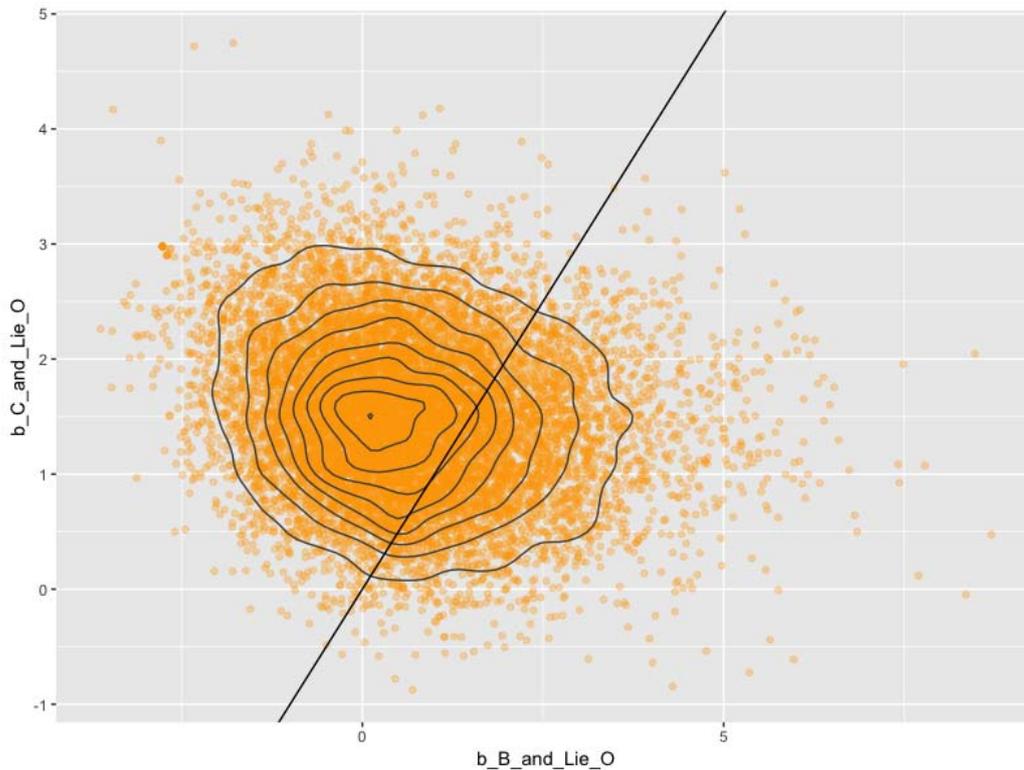

**Figure 15. Pair parameters comparison b_B_and_Lie_O and b_C_and_Lie_O**

Buddhism and Confucianism assume the same tendency in terms of correlation with lying behaviors and are distributed around the mean value. However, the coefficient of Buddhism is small, and the distribution lies between negative and positive. Thus, the results show Buddhism is not tolerant toward lying. On the contrary, Confucianism elements have a strong positive coefficient, the entire 95% CI being positive.

Next, we consider Buddhism and Taoism in their relationship with lying behaviors, using the bayesvl density plot function:

```
> bvl_plotDensity2d(model, "b_B_and_Lie_O", "b_T_and_Lie_O",
color_scheme = "orange")
```

We will also examine the correlation visually, in Figure 16 below:

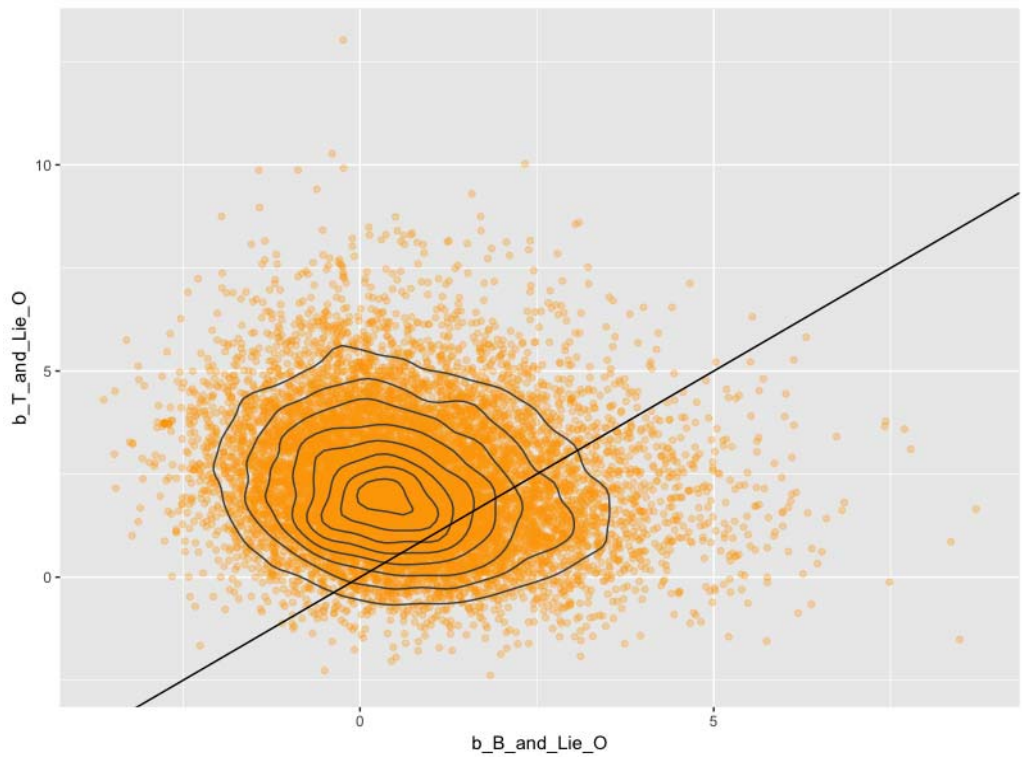

**Figure 16. Pair parameters comparison b_B_and_Lie_O and b_T_and_Lie_O**

Similar to Confucianism, Taoism also predicts lying behaviors much stronger than Buddhism does. The last pair to be compared is Confucianism and Taoism:

```
> bvl_plotDensity2d(model, "b_C_and_Lie_O", "b_T_and_Lie_O",
color_scheme = "blue")
```

A graphic representation of the result can be found in Figure 17.

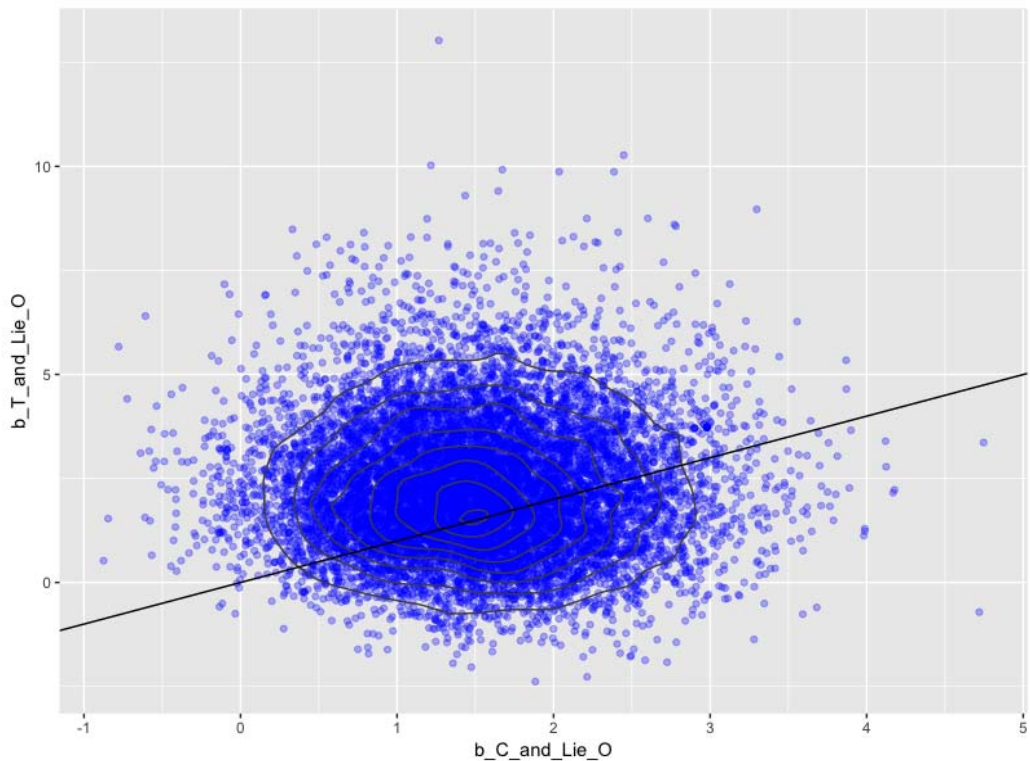

**Figure 17. Pair parameters comparison b_C_and_Lie_O and b_T_and_Lie_O**

In their relations to lying behavior of main characters, Taoism and Confucianism are quite similar and both center in the positive. One crucial characteristic of Confucianism should be kept in mind when interpreting these results: Confucianism boasts teachings related to ambitions of becoming feudal officials, the latter often being an end that justifies all means – including behaviors such as lying.

***Comparing situations with and without intervention:***

The code for tests on variable Outcome ("O") when there are and are not external elements intervening in the story yield two distribution graphs **y_rep** as follows.

```
> bvl_plotTest(model, "O", "Int1_or_Int2_1")
> bvl_plotTest(model, "O", "Int1_or_Int2_2")
```

The following pair of graphs in Figure 19 corresponds to the values of **Int1_or_Int2** in the above commands.

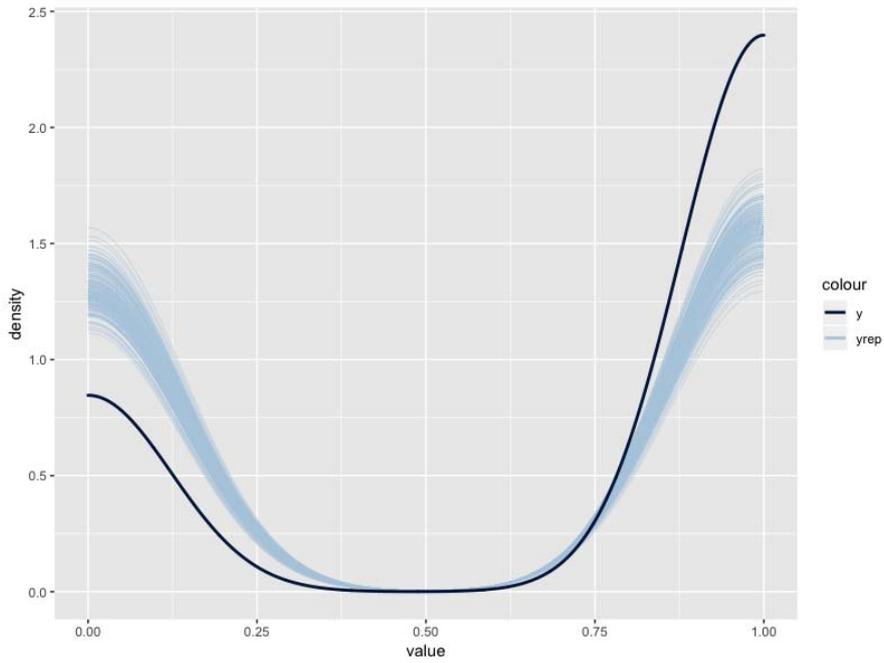

**Figure 18. Graphic representation of the relationship with the outcome variable when Int1_or_Int2=1**

Figure 18 plots the graphic representation of the relationship with outcome when **Int1_or_Int2=1**; Figure 19 when **Int1_or_Int2=2**:

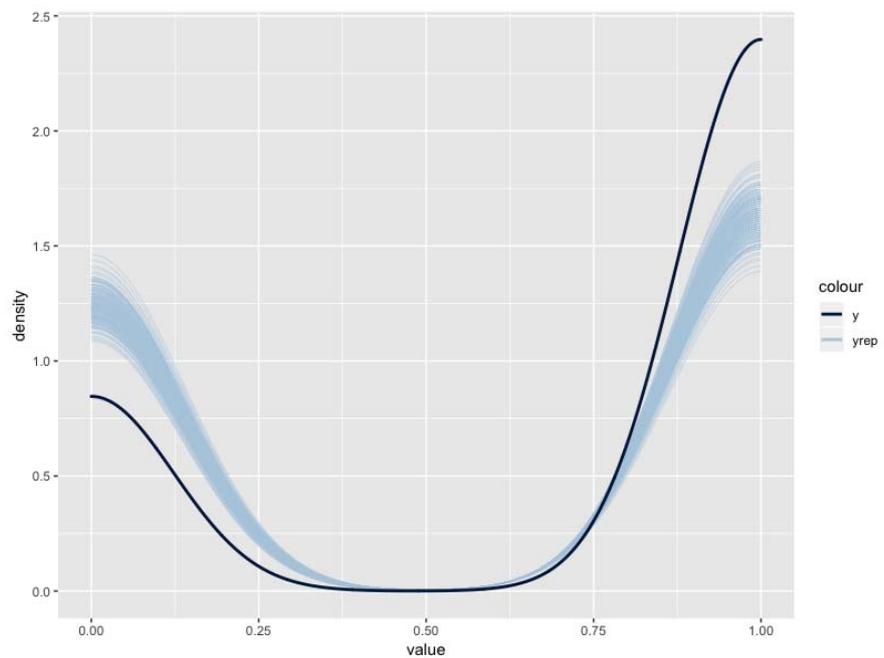

**Figure 19. Graphic representation of the relationship when Int1_or_Int2=2**

The two graphs appear rather similar, meaning that the ending of a story does not depend significantly on the factor of external intervention.

In addition, the respective distributions of coefficients **a_Int1_or_Int2[1]** and **a_Int1_or_Int2[2]** have the same curve pattern, suggesting that when there are external interventions, improvements in main character behaviors compared to when there are no interventions are still negligible. For this examination, Figures 20 and 21 are provided below.

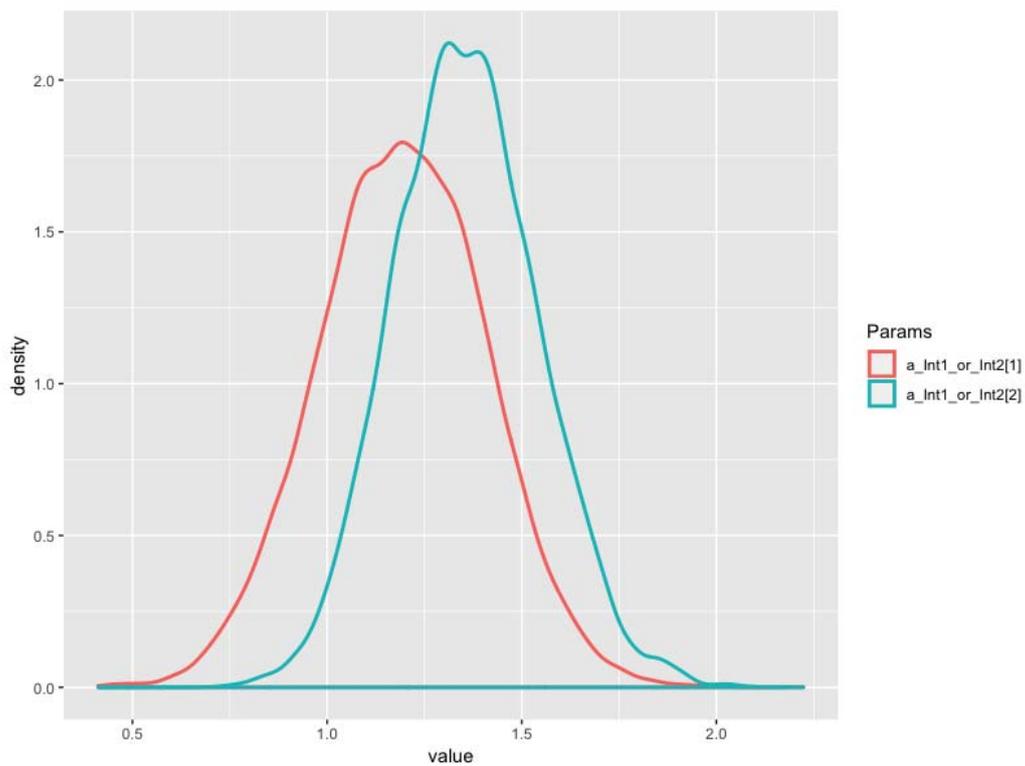

**Figure 20: Distribution of a_Int1_or_Int2[1] and a_Int1_or_Int2[2]**

The relationship between changes of coefficients **a_Int1_or_Int2[1]** and **a_Int1_or_Int2[2]** is presented in Figure 21.

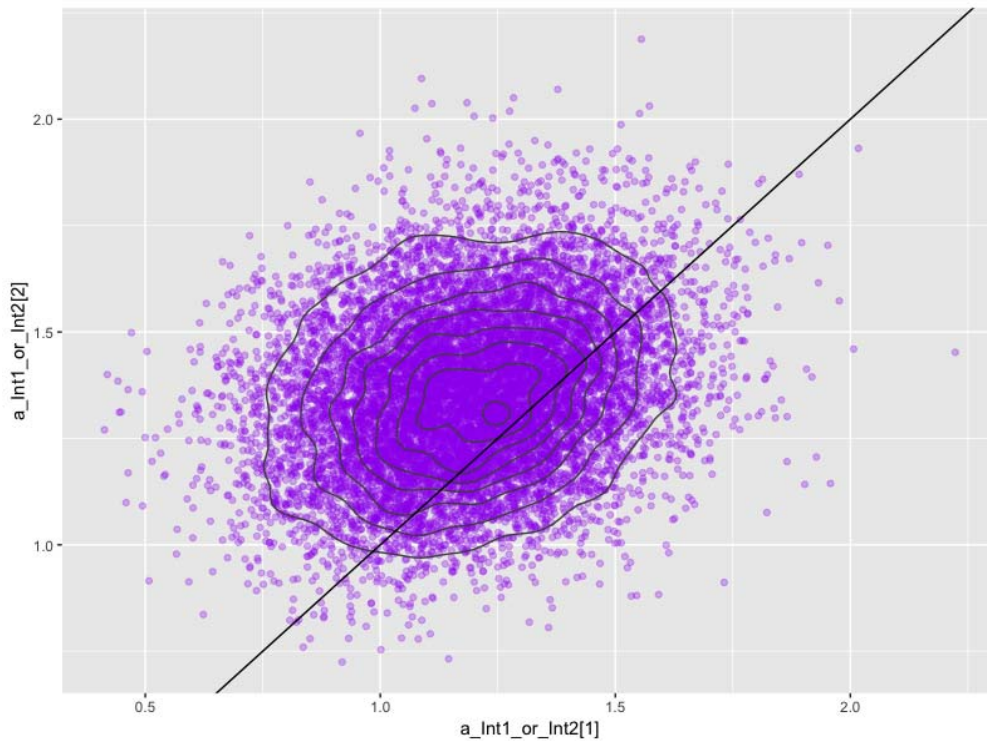

**Figure 21. The relationship between a_Int1_or_Int2[1] and a_Int1_or_Int2[2]**

Figure 21 also shows that convergent values have rather even distributions, all positive, and are mostly distribute around a determined interval of values.

**Discussion**

*Technical implications*

The use of Bayesian statistics has enabled us to efficiently analyze the complicated relationships among different variables—such as the religious teachings/values and the acts of lying and violence—when the amount of data is limited. Studies on Vietnamese folklore, though insightful in their own rights, have remained qualitative and touched on some themes such as femininity-masculinity (Do & Brennan, 2015; Nguyen, 2002), folk medical practices (Du, 1980), or psychological mindedness (Nguyễn, Foulks, & Carlin, 1991). Here, the computational method suggests there is a fertile ground for quantitative research on culture, religion, ethics, sociology, and many other social science disciplines. This direction could supplement the shortcomings that research on human neurons and brains often faced—such as making claims at the societal level.

*Sociological implications*

One of the most striking results of this study is how, in stories with Buddhist teachings, the acts of lying and/or violence are often linked to a positive outcome. This finding brings us back to the opening Buddhist story on the monk who turns into coucal and the murdering robber who gets salvation. It also resonates with what that Houben and van Kooij (1999) have once pointed out, i.e., the prevalence of violence and the rationalization of violence in South Asian folklore, particularly in stories influenced by Buddhism and Hinduism.

To make sense of the underlying mechanism behind these statistical patterns, we suggest examining which values in the *Three Teachings* are most observed by the laypeople. The tolerance for violence can be explained by the emphasis on *karma* in Buddhism – a concept might be loosely interpreted as "an eye for an eye," whereas the propensity for lying is attributable to the need to (i) preserve social order and being loyal and pious toward one's King and kins in Confucianism, and (ii) pursue spontaneity, and harmony with nature in Taoism. Historically speaking, Confucianism first started as a remedy to the chaotic, violent time of the Spring-Autumn Warring States, and Taoism was a response to the rigid way of life of the Confucianist, hence the retreat to nature and spontaneity. As these Chinese teachings prioritize social order, understandably they would be antithetical to violence and more acceptant to lying as the price in practice.

Given that all the stories were passed down through the oral tradition, they presumably do not evoke the true ideals of the three religions, but instead, reflect the psychology and understanding of the people at the time. Future studies could replicate the method beyond the folklore realm to see if these observations hold. What is highlighted here is a glaring double standard in the interpretation and practice of the three teachings: the very virtuous outcomes being preached, whether that be compassion and meditation in Buddhism, societal order in Confucianism, or natural harmony in Taoism, appear to accommodate two universal vices—violence in Buddhism and lying in the latter two. Attempts to make sense of contradictory human behaviors have pointed out the role of cognition in belief maintenance and motivated reasoning in discounting counterargument (Bersoff, 1999; Kaplan, Gimbel, & Harris, 2016). When it comes to religion, an individual's tolerance of contradictory religious teachings is not due to lower rationality standards but rather due to how such teachings fit the "inference machinery" in a plausible manner (Boyer, 2001). This study takes a step further by showing how such action could even result in desirable outcomes, even if the findings may be limited to the folkloristic realm. What is clearly troubling is the promotion of the ends-justify-the-means mentality when the acceptance of values counter to one's beliefs is correlated with positive outcomes.

**Conclusion**

The present study, through the Bayesian network analysis of 307 Vietnamese folktales, has reached two notable conclusions. First, folktale characters who commit either lying or violence face negative outcomes in general, but there is a mixed result when the religious values are taken into consideration.

In particular, lying for characters associated with Confucianism or Taoism, not for Buddhism, tends to bring about a positive outcome ($β_{T\_and\_Lie\_O}$= 2.23; $β_{C\_and\_Lie\_O}$= 1.47). Second, the typical character associated with Buddhism, who commits violence tends to have a happy ending ($β_{B\_and\_Viol\_O}$= 2.55). These findings raise questions about the double standard when people interpret and practice the teachings of Buddhism, Confucianism, and Taoism. Although the three Teachings preach that one ought to cultivate moral characters, it seems in folktales as well as in life, followers of such teachings are no exception to the two universal vices, lying, and violence. Moreover, in certain cases, these vices may bring about by positive outcomes, as detected by the statistical technique deployed in this study. Such contradiction casts doubt on the meanings of religious teachings and at the same time, calls into questions the complexity of human decision-making, especially beyond the folklore realm.

**Data deposition:**

The dataset is available from the Github (https://github.com/sshpa/bayesvl/tree/master/data), and CRAN 'bayesvl' R package (https://cran.r-project.org/package=bayesvl). The executable computer code for this paper is provided in the available on https://github.com/sshpa/bayesvl/tree/master/examples, file: simulation_example.R.

**References**

Abello, J., Broadwell, P., & Tangherlini, T. R. (2012). Computational folkloristics. *Communications of the ACM, 55(7)*(7), 60-70.

Albrecht, S. L., Chadwick, B. A., & Alcorn, D. S. (1977). Religiosity and deviance: Application of an attitude-behavior contingent consistency model. *Journal for the Scientific Study of Religion, 16*(3), 263-274. doi:10.2307/1385697

Alcantud-Diaz, M. (2010). Violence in the Brothers Grimm's fairy tales: A corpus-based approach. *Revista Alicantina de Estudios Ingleses*, 173-185. doi:10.14198/raei.2010.23.10

Alcantud-Diaz, M. (2014). "Kill her, and bring me back her heart as a Token!" identity, power, and violence in the Grimm's fairy tales collection. *International Journal on Studies in English Language and Literature (IJSELL), 2*, 84-99.

Atran, S. (2016). The devoted actor: Unconditional commitment and intractable conflict across cultures. *Current Anthropology, 57*(S13), S192-S203. doi:10.1086/685495

Benda, B. B. (2002). Religion and violent offenders in boot camp: A structural equation model. *Journal of Research in Crime and Delinquency, 39*(1), 91-121. doi:10.1177/002242780203900104

Bersoff, D. M. (1999). Why good people sometimes do bad things: Motivated reasoning and unethical behavior. *Personality and Social Psychology Bulletin, 25*(1), 28-39. doi:10.1177/0146167299025001003

Blogowska, J., Lambert, C., & Saroglou, V. (2013). Religious prosociality and aggression: It's real. *Journal for the Scientific Study of Religion, 52*(3), 524-536. doi:10.1111/jssr.12048

Bortolini, E., Pagani, L., Crema, E. R., Sarno, S., Barbieri, C., Boattini, A., Sazzini, M., Da Silva, S. G., Martini, G., & Metspalu, M. (2017a). Inferring patterns of folktale diffusion using genomic data. *Proceedings of the National Academy of Sciences, 114*(34), 9140-9145.

Bortolini, E., Pagani, L., Crema, E. R., Sarno, S., Barbieri, C., Boattini, A., Sazzini, M., da Silva, S. G., Martini, G., & Metspalu, M. (2017b). Reply to d'Huy et al.: Navigating biases and charting new ground in the cultural diffusion of folktales. *Proceedings of the National Academy of Sciences, 114*(41), E8556-E8556.

Boyer, P. (2001). *Religion explained: The evolutionary origins of religious thought*. New York: Basic Books.

Brooks, S. P., & Gelman, A. (1998). General methods for monitoring convergence of iterative simulations. *Journal of Computational and Graphical Statistics, 7*(4), 434-455.

Bruggeman, E. L., & Hart, K. J. (1996). Cheating, lying, and moral reasoning by religious and secular high school students. *The Journal of Educational Research, 89*(6), 340-344. doi:10.1080/00220671.1996.9941337

Casadio, G. (2016). *Historicizing and translating religion*. Oxford: Oxford University Press.

Chima, A., & Helen, G. (2015). Evil and superstition in Sub-Saharan Africa: Religious infanticide and filicide perceiving evil. In D. Farnell, R. Noiva, & K. Smith (Eds.), *Perceiving Evil: Evil Women and the Feminine* (pp. 33-48). Leiden, The Netherlands: Brill.

Cleary, J. C. (1991). Buddhism and popular religion in medieval Vietnam. *Journal of the American Academy of Religion, 59*(1), 93-118.

Cochran, J. K., & Akers, R. L. (1989). Beyond hellfire: An exploration of the variable effects of religiosity on adolescent marijuana and alcohol use. *Journal of Research in Crime and Delinquency, 26*(3), 198-225. doi:10.1177/0022427889026003002

Corcoran, K. E., Pettinicchio, D., & Robbins, B. (2012). Religion and the acceptability of white-collar crime: A cross-national analysis. *Journal for the Scientific Study of Religion, 51*(3), 542-567. doi:10.1111/j.1468-5906.2012.01669.x

Coupland, J., & Jaworski, A. (2003). Transgression and intimacy in recreational talk narratives. *Research on Language and Social Interaction, 36*(1), 85-106. doi:10.1207/S15327973RLSI3601_5

d'Huy, J., Le Quellec, J.-L., Berezkin, Y., Lajoye, P., & Uther, H.-J. (2017). Studying folktale diffusion needs an unbiased dataset. *Proceedings of the National Academy of Sciences*, 201714884.

Do, V. H. T., & Brennan, M. (2015). Complexities of Vietnamese femininities: A resource for rethinking women's university leadership practices. *Gender and Education, 27*(3), 273-287.

Dogra, S. (2018). Folklore and Computers: The Oral and the Digital in Computational Folkloristics. *AN INTERNATIONAL JOURNAL OF HUMANITIES AND SOCIAL SCIENCES, 6*(1), 22-27.

Doré, B., & Bolger, N. (2018). Population-and individual-level changes in life satisfaction surrounding major life stressors. *Social Psychological and Personality Science, 9*(7), 875-884.

Downey, A. B. (2012). *Think Bayes - Bayesian statistics made simple*. Massachusetts: Green Tea Press.

Du, J. N. (1980). Pseudobattered child syndrome in Vietnamese immigrant children. *Canadian Medical Association Journal, 122*(4), 394.

Durkheim, E. (1897). *Le suicide*. Paris: Presses Électroniques de France.

Editorial. (2017). Promoting reproducibility with registered reports. *Nature Human Behaviour, 1*(1), 0034. doi:10.1038/s41562-016-0034

Evans, T. D., Cullen, F. T., Dunaway, R. G., & Burton Jr, V. S. (1995). Religion and crime reexamined: The impact of religion, secular controls, and social ecology on adult criminality. *Criminology, 33*(2), 195-224. doi:10.1111/j.1745-9125.1995.tb01176.x

Festinger, L. (1962). *A theory of cognitive dissonance* (Vol. 2): Stanford University Press.

Gelman, A. (2008). Objections to Bayesian statistics. *Bayesian Analysis, 3*, 445-450.

Gill, J. (2002). *Bayesian methods: A social and behavioral sciences approach*: Chapman and Hall/CRC.

Graham, J., Meindl, P., Beall, E., Johnson, K. M., & Zhang, L. (2016). Cultural differences in moral judgment and behavior, across and within societies. *Current Opinion in Psychology, 8*, 125-130. doi:https://doi.org/10.1016/j.copsyc.2015.09.007

Haar, B. t. (2005). *Telling Stories*. Leiden, The Netherlands: Brill.

Haidt, J., & Graham, J. (2007). When morality opposes justice: Conservatives have moral intuitions that liberals may not recognize. *Social Justice Research, 20*(1), 98-116. doi:10.1007/s11211-007-0034-z

Haidt, J., & Joseph, C. (2004). Intuitive ethics: How innately prepared intuitions generate culturally variable virtues. *Daedalus, 133*(4), 55-66.

Henrich, J., Bauer, M., Cassar, A., Chytilová, J., & Purzycki, B. G. (2019). War increases religiosity. *Nature Human Behaviour, 3*(2), 129-135. doi:10.1038/s41562-018-0512-3

Henrich, J., Heine, S. J., & Norenzayan, A. (2010). The weirdest people in the world? *Behavioral and Brain Sciences, 33*(2-3), 61-83. doi:10.1017/S0140525X0999152X

Hirschi, T., & Stark, R. (1969). Hellfire and delinquency. *Social Problems, 17*(2), 202-213. doi:10.2307/799866

Holman, D., & Walker, A. (2018). Social Quality and Health: Examining Individual and Neighbourhood Contextual Effects Using a Multilevel Modelling Approach. *Social Indicators Research, 138*(1), 245-270. doi:10.1007/s11205-017-1640-2

Houben, J. E. M., & van Kooij, K. R. (Eds.). (1999). *Violence denied: violence, non-violence, and the rationalization of violence in South Asian cultural history* (Vol. 16): Brill.

Jackman, S. (2000). Estimation and inference are missing data problems: Unifying social science statistics via Bayesian simulation. *Political Analysis, 8*(4), 307-332. doi:10.1093/oxfordjournals.pan.a029818

Jackman, S. (2009). *Bayesian analysis for the social sciences* (Vol. 846): John Wiley & Sons.

Kaplan, J. T., Gimbel, S. I., & Harris, S. (2016). Neural correlates of maintaining one's political beliefs in the face of counterevidence. *Scientific Reports, 6*, 39589. doi:10.1038/srep39589

_

https://www.nature.com/articles/srep39589#supplementary-information

Kendall, L. (2011). Gods, gifts, markets, and superstition: spirited consumption from Korea to Vietnam. In K. W. Endres & A. Lauser (Eds.), *Engaging the Spirit World: Popular Beliefs and Practices in Modern Southeast Asia* (pp. 103-120).

Kim, S. J., Song, A., Lee, G.-L., & Bach, A. (2018). Using animated folktales to teach cultural values: A case study with Korean-American bilingual kindergartners. *Journal of Research in Childhood Education, 32*(3), 295-309. doi:10.1080/02568543.2018.1464528

Kruschke, J. K. (2015). *Doing Bayesian data analysis: A tutorial with R, JAGS, and Stan* (2nd ed.). London, UK: Elsevier.

La, V. P., & Vuong, Q. H. (2019). bayesvl: Visually learning the graphical structure of Bayesian networks and performing MCMC with 'Stan'>. *The Comprehensive R Archive Network. Available from: https://cran.r-project.org/package=bayesvl*. Retrieved June 1, 2019, from https://cran.r-project.org/package=bayesvl

MacDonald, M. R. (2013). Sharing SE Asian Folktales for Character Education. *IBBY Asia Oceania Conference Bali.* Retrieved September 16, 2019, from http://www.ibby.org/fileadmin/user_upload/04-Margaret_MacDonald-Sharing_SE_Asian_Folktales_for_Character_Education.pdf

MacDonald, M. R., & Vathanaprida, S. (1994). *Thai Tales: Folktales of Thailand (World Folklore Series)*. Exeter, UK: Libraries Unlimited.

Malakoff, D. (1999). Bayes offers a 'new' way to make sense of numbers. *Science, 286*(5444), 1460-1464. doi:10.1126/science.286.5444.1460 %J Science

McElreath, R. (2016). *Statistical rethinking: A Bayesian course with examples in R and Stan*. Boca Raton, FL: CRC Press.

McKay, R., & Whitehouse, H. (2015). Religion and morality. *Psychological Bulletin, 141*(2), 447-473. doi:10.1037/a0038455

Meehan, B. (1994). Son of Cain or Son of Sam? The monster as the serial killer in Beowulf. *Connecticut Review, 16*(2), 1-7.

Mensah, C., & Azila-Gbettor, E. M. (2018). Religiosity and students' examination cheating: evidence from Ghana. *International Journal of Educational Management, 32*(6), 1156-1172. doi:10.1108/IJEM-07-2017-0165

Michalopoulos, S., & Xue, M. M. (2019). Folklore. *National Bureau of Economic Research Working Paper Series, No. 25430*. doi:10.3386/w25430

Møller, A. P., Morelli, F., & Tryjanowski, P. (2017). Cuckoo folklore and human well-being: Cuckoo calls predict how long farmers live. *Ecological Indicators, 72*, 766-768. doi:https://doi.org/10.1016/j.ecolind.2016.09.006

Muth, C., Oravecz, Z., & Gabry, J. (2018). User-friendly Bayesian regression modeling: A tutorial with rstanarm and shinystan. *Quantitative Methods for Psychology, 14*(2), 99-119.

Nguyen, D., Trieschnigg, D., & Theune, M. (2013). *Folktale classification using learning to rank.* Paper presented at the European Conference on Information Retrieval.

Nguyen, L. (2014). *Viet Nam Phat Giao Su Luan [On the history of Buddhism in Vietnam]*. Hanoi, Vietnam: NXB Van Hoc.

Nguyen, M. D. (1985). Culture Shock - A Review of Vietnamese Culture and Its Concepts of Health and Disease. *Cross-cultural Medicine, 142*(3), 409-412.

Nguyễn, N., Foulks, E. F., & Carlin, K. (1991). Proverbs as psychological interpretations among Vietnamese. *Asian Folklore Studies, 50*(2), 311-318. doi:10.2307/1178388

Nguyen, N. H. (1998). The Confucian incursion into Vietnam. In W. H. Slote & G. A. D. Vos (Eds.), *Confucianism and the Family: A Study of Indo-Tibetan Scholasticism* (pp. 91-104).

Nguyen, T. A. (1993). Buddhism and Vietnamese Society Throughout History. S*outheast Asia Research, 1*(1), 98-114.

Nguyen, T. T. (2008). *History of Buddhism in Vietnam*. Washington D.C.: Council for Research in Values & Philosophy.

Nguyen, V. K. (2002). Rethinking the status of Vietnamese women in folklore and oral history. In P. Brocheux & G. L. Bousquet (Eds.), *Viet-Nam Expose-French Scholarship on Twentieth-Century Vietnamese Society*. Ann Arbor: University of Michigan Press.

Nikolić, D., & Bakarić, N. (2016). What makes our tongues twist?: Computational analysis of Croatian tongue-twisters. *The Journal of American Folklore, 129*(511), 43-54. doi:10.5406/jamerfolk.129.511.0043

Perrin, R. D. (2000). Religiosity and honesty: Continuing the search for the consequential dimension. *Review of Religious Research, 41*(4), 534-544. doi:10.2307/3512319

Purzycki, B. G., & Gibson, K. (2011). Religion and violence: an anthropological study on religious belief and violent behavior. *Skeptic, 16*(2), 22-27.

Rettinger, D. A., & Jordan, A. E. (2005). The relations among religion, motivation, and college cheating: A natural experiment. *Ethics & Behavior, 15*(2), 107-129. doi:10.1207/s15327019eb1502_2

Rohrbaugh, J., & Jessor, R. (1975). Religiosity in youth: A personal control against deviant behavior1. *Journal of Personality, 43*(1), 136-155. doi:10.1111/j.1467-6494.1975.tb00577.x

Rohrbaugh, J., & Jessor, R. (2017). Religiosity: A personal control against delinquency. In R. Jessor (Ed.), *Problem Behavior Theory and Adolescent Health: The Collected Works of Richard Jessor, Volume 2* (pp. 393-409). Cham: Springer International Publishing.

Sandberg, S. (2014). What can "lies" tell us about life? Notes towards a framework of narrative criminology. In H. Copes (Ed.), *Advancing Qualitative Methods in Criminology and Criminal Justice* (pp. 68-86): Routledge.

Sandberg, S., Tutenges, S., & Copes, H. (2015). Stories of violence: A narrative criminological study of ambiguity. *The British Journal of Criminology, 55*(6), 1168-1186. doi:10.1093/bjc/azv032

Scales, J. A., & Snieder, R. (1997). To Bayes or not to Bayes? *Geopolitics, 62*(4), 1045-1046.

Scutari, M. (2010). Learning Bayesian Networks with the bnlearn R Package. *Journal of Statistical Software, 35*(3), 1-22.

Spiegelhalter, D. (2019). *The Art of Statistics: Learning from data*. London, UK: Penguin Random House UK.

Tangherlini, T. R. (2013). The Folklore Macroscope. *Western Folklore, 72*(1), 7-27.

Tehrani, J. J., & d'Huy, J. (2017). Phylogenetics Meets Folklore: Bioinformatics Approaches to the Study of International Folktales. In R. Kenna, M. MacCarron, & P. MacCarron (Eds.), *Maths Meets Myths: Quantitative Approaches to Ancient Narratives. Understanding Complex Systems* (pp. 91-114). Cham: Springer.

Thao, H. T. P., & Vuong, Q.-H. (2015). A Merton model of credit risk with jumps. *Journal of Statistics Applications and Probability Letters, 2*(2), 97-103.

Thenmozhi, K., Anusuya, N., Ajmal Ali, M., Jamuna, S., Karthika, K., Venkatachalapathi, A., Al-Hemaid, F. M., Farah, M. A., & Paulsamy, S. (2018). Pharmacological credence of the folklore use of Bauhinia malabarica in the management of jaundice. *Saudi Journal of Biological Sciences, 25*(1), 22-26. doi:https://doi.org/10.1016/j.sjbs.2017.08.001

Tian, X. (2014). Rumor and Secret Space: Organ-Snatching Tales and Medical Missions in Nineteenth-Century China. *Modern China, 41*(2), 197-236. doi:10.1177/0097700414525614

Tittle, C. R., & Welch, M. R. (1983). Religiosity and deviance: Toward a contingency theory of constraining effects. *Social Forces, 61*(3), 653-682. doi:10.1093/sf/61.3.653

Toan-Anh. (2005). *Nep cu: Tin nguong in Vietnam (Quyen Thuong) [Old habits: Religious beliefs in Vietnam (Vol. I)]*. Hanoi: Nha Xuat Ban Tre.

Towhidloo, Y., & Shairi, H. R. (2017). A study of making prostheses in discourse: Why does narrative lie make prostheses? *Research in Contemporary Word Literature, 22*(1 #A00119), 269-286. doi:10.22059/jor.2017.62714

Tran, A. Q. (2017). *Gods, Heroes, and Ancestors: An Interreligious Encounter in Eighteenth-century Vietnam*: Oxford University Press.

Vallerand, R. J. (1997). Toward a hierarchical model of intrinsic and extrinsic motivation. In M. P. Zanna (Ed.), *Advances in Experimental Social Psychology* (Vol. 29). US: Academic Press.

Victor, J. (1990). Satanic cult rumors as contemporary legend. *Western Folklore, 49*(1), 51-81. doi:10.2307/1499482

Vuong, Q.-H., Bui, Q.-K., La, V.-P., Vuong, T.-T., Ho, M.-T., Nguyen, H.-K. T., Nguyen, H.-N., Nghiem, K.-C. P., & Ho, M.-T. (2019a). Cultural evolution in Vietnam's early 20th century: A Bayesian networks analysis of Hanoi Franco-Chinese house designs. *Social Sciences & Humanities Open, 1*(1), 100001. doi:https://doi.org/10.1016/j.ssaho.2019.100001

Vuong, Q.-H., Bui, Q.-K., La, V.-P., Vuong, T.-T., Nguyen, V.-H. T., Ho, M.-T., Nguyen, H.-K. T., & Ho, M.-T. (2018). Cultural additivity: behavioural insights from the interaction of Confucianism, Buddhism and Taoism in folktales. *Palgrave Communications, 4*(1), 143. doi:10.1057/s41599-018-0189-2

Vuong, Q. H., Ho, M.-T., & La, V. P. (2019b). 'Stargazing' and p-hacking behaviours in social sciences: some insights from a developing country. *European Science Editing, 45*(2), 54-55.

Wang, X., & Jang, J. S. (2018). The effects of provincial and individual religiosity on deviance in China: A multilevel modeling test of the moral community thesis. *Religions, 9*(7), 202. doi:10.3390/rel9070202

Welch, M. R., Tittle, C. R., & Grasmick, H. G. (2006). Christian religiosity, self-control and social conformity. *Social Forces, 84*(3), 1605-1623. doi:10.1353/sof.2006.0075

Xu, Y. Z. (2002). A thesis on the spreading and influence of Taoism in Vietnam. *Journal of Historical Science, 7*, 015.

Yun, I., & Lee, J. (2016). The relationship between religiosity and deviance among adolescents in a religiously pluralistic society. *International Journal of Offender Therapy and Comparative Criminology, 61*(15), 1739-1759. doi:10.1177/0306624X16657622